\documentclass[leqno]{article}

\newcommand{\titel}
{Three Datatype Defining Rewrite Systems for Datatypes of Integers  
each extending a Datatype of Naturals}

\renewcommand{\equiv}{=}
\usepackage{stmaryrd,amssymb,amsmath,amsthm,url}

\usepackage{environ}
\makeatletter
\NewEnviron{Lalign}{\tagsleft@true\begin{align}\BODY\end{align}}
\makeatother

\usepackage{environ}
\makeatletter
\NewEnviron{Ralign}{\tagsleft@false\begin{align}\BODY\end{align}}
\makeatother

\theoremstyle{definition}

\newcommand{\Nat}{{\mathbb N}}
\newcommand{\NF}{\mathit{N}}
\newcommand{\Int}{{\mathbb Z}}
\newcommand{\PNF}{{\NF^+}}
\newcommand{\NNF}{{\NF^-}}

\newcommand{\app}[1]{\hspace{.4mm}{:_{\hspace{.2mm}#1}}\hspace{.4mm}}
\newcommand{\apu}{\app u}
\newcommand{\apue}{\apu1}
\newcommand{\apb}{\app b}
\newcommand{\apbz}{\apb 0}
\newcommand{\apbu}{\apb 1}
\newcommand{\apbo}{\apbu}
\newcommand{\apd}{\app d}
\newcommand{\apt}{\app t}

\newcommand{\conca}[1]{\hspace{1mm}\hat{\raisebox{-,4ex}{\footnotesize{\it #1}}}\hspace{1mm}}
\newcommand{\concu}{\conca u}
\newcommand{\concb}{\conca b}
\newcommand{\concd}{\conca d}

\begin{document}

\title{\titel\thanks{Version 4: 
All DDRSes defined in Section~\ref{sec:2} are proven ground-complete (Appendix~\ref{app:A}). 
In Section~\ref{sec:3}, the DDRS for $\Int_{ut}$ in Table~\ref{tab:intuaview} now contains only 
sixteen equations and is proven ground-complete; the
DDRS for $\Int_{bt}$ in Table~\ref{natbview} has one more equation (\ref{bi22}) and is 
proven ground-complete; the DDRSes for $\Nat_{dt}$ (Table~\ref{tab:natdecview}) and $\Int_{dt}$ 
(Table~\ref{tab:intdecview}) are those that are proven ground-complete in~\cite{Kamp16}.
In Appendix~\ref{app:C}, corrected versions of the DDRSes for $\Nat_{u'}$ 
and $\Int_{u'}$ are proven ground-complete.
}}

\author{
	Jan A.\ Bergstra ~and~ Alban Ponse
	\\[2mm]
  {\small
	  Informatics Institute,}\\
	  {\small section Theory of Computer Science}, \\
	  {\small University of Amsterdam}\\[2mm]
	 {\small \url{https://staff.fnwi.uva.nl/j.a.bergstra/}\quad\url{https://staff.fnwi.uva.nl/a.ponse/}
	}
}
\date{}

\maketitle

\thispagestyle{empty}

\begin{abstract}
Integer arithmetic is specified according to three views: unary, binary, and decimal notation.
The binary and decimal view have as their characteristic that each normal form resembles common
number notation, that is, either a digit, or a string of digits without leading zero, or the 
negated versions of the latter. 
The unary view comprises a specification of integer
arithmetic based on 0, successor function $S$, and predecessor function, with negative
normal forms $-S^i(0)$.
Integer arithmetic in binary and decimal notation is based on (postfix) digit append functions.
For each view we define a ground-confluent and terminating datatype defining rewrite system (DDRS), 
and in each case the resulting datatype is a canonical term algebra that extends a corresponding
canonical term algebra for natural numbers. 

Then, for each view, we consider an 
alternative DDRS based on tree constructors that yield comparable normal forms, which for
that binary and decimal view admits expressions that are algorithmically more involved. 
These DDRSes are incorporated because they are closer to existing literature.
For these DDRSes we also provide ground-completeness results. 

Finally, we define a DDRS for the ring of Integers (comprising fifteen rewrite rules) and
prove its ground-completeness.
\\[2mm]
\end{abstract}

\vfill

{\tableofcontents}

\section{Introduction}
We specify integer arithmetic according to three different ``views'': unary, binary, and decimal 
notation. This work is based on the specifications for natural numbers from \cite{Bergstra2014n}
and we follow the same strategy to develop these different views. 
Each of the specifications we provide is a so-called
DDRS (datatype defining rewrite system) and consists of a number of equations that define
a {term rewriting system} when interpreting the equations from left-to-right. 
A DDRS must be ground-complete, that is, strongly terminating and ground-confluent; 
for some general information on term rewriting systems see e.g.~\cite{Terese}.

This paper constitutes a further stage in the development of a family of arithmetical 
datatypes with corresponding specifications. 
The resulting specifications (DDRSes) incorporate different views on the same abstract datatype
(ADT), where an ADT may be understood as the isomorphism class of its instantiations 
which are concrete datatypes. 
The datatypes considered in~\cite{Bergstra2014n} are so-called canonical term algebras
which means that carriers are non-empty sets of closed terms which are closed under taking subterms.

The \emph{unary view} provides a term rewriting system where terms in unary notation serve as normal 
forms. 
The unary view also provides a semantic specification of binary notation, of decimal notation, 
and of hexadecimal notation. These three positional notations were modified in~\cite{Bergstra2014n}
with respect to conventional notations in such a way that syntactic confusion between these 
notations cannot arise. 
In this paper, the \emph{hexadecimal view} is left out as that seems to be an unusual viewpoint 
for integer arithmetic.

It seems to be the case that for the unary view the specification of the integers 
(see Table~\ref{tab:intunview}) is entirely
adequate, whereas all subsequent specifications for binary view and decimal view may provide no more
than a formalization of a topic which must be somehow understood before taking notice of that 
same formalization. It remains to be seen to what extent the first DDRS for the unary case may serve 
exactly that expository purpose.

The strategy of this work is somewhat complicated: on the one hand we 
look for specifications that may genuinely be considered introductory, 
that is, descriptions that can be used to construct the datatype at hand 
for the first time in the mind of a person.
On the other hand awareness of the datatype in focus may be needed to 
produce an assessment of the degree of success achieved in the direction of the first objective.

In the remainder of this section we discuss some preliminaries.
In Section~\ref{sec:2} we provide for each view two DDRSes, one for natural number arithmetic,
and one for integer arithmetic.
In Section~\ref{sec:3} we consider for each view  
alternative DDRSes based on tree constructors that yield comparable normal forms
and are closer to existing literature.
In Section~\ref{sec:4} we finish the paper with some concluding remarks, and we define
a DDRS for arithmetic in the ``language of rings'' that is of some theoretical interest 
(it is used in~\cite{BP16}), while its relatively small ground-confluence proof paves the way
to the more elaborated ground-confluence proofs for the DDRSes defined in Sections~\ref{sec:2} 
and~\ref{sec:3} that are recorded in 
Appendix~\ref{app:A} and Appendix~\ref{app:B}, respectively. 
Finally, in Appendix~\ref{app:C} we 
discuss two very simple DDRSes, the first one of which establishes an alternative model for 
natural number arithmetic in unary view and the second one its generalization to 
integer arithmetic.

\subsection{Digits and rewrite rules in equational form}
Digits are elements of the set $D=\{0, 1, 2, 3, 4, 5, 6, 7, 8, 9\}$, ordered in the common way:
\[0 < 1 < 2 < 3 <  4 < 5 <  6 < 7 < 8 < 9.\]

For the digits $0, 1, \dots, 8$ we denote with $i^{\prime}$ the successor digit of $i$ 
in the given enumeration.
In Table~\ref{fig:enumdig} the successor notation on digits is specified as a transformation 
of syntax, and we adopt this notation throughout the paper.

\begin{table}
\centering
\hrule
\begin{align*}
0^{\prime} &\equiv 1&3^{\prime} &\equiv 4&6^{\prime} &\equiv 7\\
1^{\prime} &\equiv 2&4^{\prime} &\equiv 5&7^{\prime} &\equiv 8\\
2^{\prime} &\equiv 3&5^{\prime} &\equiv 6&8^{\prime} &\equiv 9
\end{align*}
\hrule
\caption{Enumeration and successor notation of digits of type $\Int$}
\label{fig:enumdig}
\end{table}

We will list rewrite rules in the form of equations $t=r$ to be interpreted from left-to-right, 
and we will add tags of the form
\[\text{[N$n$]}~~t=r\]
for reference, with ``N'' some name and ``$n$'' a natural number (in ordinary, decimal notation).
Furthermore, for $k,\ell \in D$ and $k < \ell$, the notation
\[\text{[N$n.i\hspace{.4mm}]_{i=k}^\ell~~t=r$}\]
represents the following $\ell - k+1$ equations:
\[\text{[N$n.k$]~~} t[k/i]=r[k/i],~\dots ,~\text{[N$n.\ell$]~~} t[\ell/i]=r[\ell/i],\]
thus with $i$ instantiated from $k$ to $\ell$. 
Occasionally, we will use this notation with two  ``digit counters'', as in
\[\text{[N$n.i.j\hspace{.4mm}]_{i,j=k}^\ell~~t=r$},
\]
for a concise representation of the following $(\ell-k+1)^2$ equations:
\[\begin{aligned}
&\text{[N$n.k.k$]~~} t[k/i][k/j]=r[k/i][k/j], ~\dots,&\text{[N$n.k.\ell$]~~} 
t[k/i][\ell/j]=r[k/i][\ell/j],\\
&\ldots,\\
&\text{[N$n.\ell.k$]~~} t[\ell/i][k/j]=r[\ell/i][k/j], 
~~\dots,&\text{[N$n.\ell.\ell$]~~} t[\ell/i][\ell/j]=r[\ell/i][\ell/j].~
\end{aligned}\]

\subsection{A signature for integers}
\label{sec:sig}
The signature $\Sigma_{\Int}$ has the following elements:
\begin{enumerate}

\item A sort $\Int$, 

\item For digits the ten  constants $0,1,2,3,4,5,6,7,8,9$, 

\item Three one-place functions 
$S,P,-:\Int \to \Int$,
``successor'', ``predecessor'', and  ``minus'', respectively,

\item Addition and multiplication (infix) 
$+,\cdot: \Int \times \Int \to \Int,$

\item 
\label{it:5}
Two one-place functions (postfix)
\(\_\apbz, ~\_\apbu: \Int \to \Int,\)
``binary append zero'' and ``binary append one'',
these functions will be used for binary 
notation,

\item 
\label{it:6}
Ten one-place functions (postfix)
\begin{align*}
&\_\apd 0, ~\_\apd 1, ~\_\apd 2, ~\_\apd 3, ~\_\apd 4, ~\_\apd 5,
~\_\apd 6, ~\_\apd 7, ~\_\apd 8, ~\_\apd 9:\Int \to \Int,
\end{align*}
``decimal append zero'', \dots,``decimal append nine'', to be used for decimal 
notation.
\end{enumerate}
We shall use the following abbreviations, where $i$ is a digit:
$S^i(t)$ stands for $i$ applications of the successor function $S$ to $t$, 
thus $S^0(t)=t$ and 
$S^{i'}(t)=S(S^i(t))$, and
$P^i(t)$ stands for $i$ applications of the predecessor function $P$ to $t$.

The ``append $<$digit name$>$'' functions defined in
items~\ref{it:5} and \ref{it:6} can be viewed as instantiations of 
more general two-place ``append'' functions, but that would require the introduction of sorts 
for bits (binary digits) and for decimal digits. 
However, we prefer to keep the signature single-sorted
and that is why we instantiate such ``digit append'' functions per digit to unary functions
and why we use postfix notation for applications of these functions. E.g., 
\[(9\apd7)\apd5\quad \text{and}\quad ((1\apbz)\apbz)\apbu\]
represent the decimal number 975, and the binary number $1001$, respectively.
For the unary view the normal forms are the classical successor terms, that is 
\[0,S(0), S(S(0)),\dots\]
and all minus instances $-(t)$ of each such nonzero normal form $t$, e.g.~$-(S(S(0)))$,
if no confusion can arise, we abbreviate $-(t)$ to $-t$, as in $-x$. 

For the binary view and for the decimal view, we provide one DDRS for each.
Normal forms are all appropriate digits, all applications of the respective 
append functions to a nonzero normal form, and all minus instances $-t$ of each 
such normal form $t$ that differs from 0.
Thus $-(((1\apbz)\apbz)\apbu)$ is an example of a normal form in binary view, and
$-((9\apd7)\apd5)$ is one in decimal view.

\section{Three DDRSes for datatypes of Integers, each extending a datatype of Naturals}
\label{sec:2}
We provide for each of the unary, binary and decimal view two DDRSes, one for natural 
number arithmetic and one for integer arithmetic.
For the binary and decimal view we consider specifications
that also employ the successor and predecessor functions. These specifications
are far more lengthy and involved, but as DDRSes their quality improves because
normal forms are smaller and are reached in fewer rewriting steps.

\subsection{Unary view}
\label{subsec:2.1}
Table~\ref{tab:natunview} provides a DDRS for the natural numbers and defines the 
canonical term algebra 
$\Nat_{ubd}$, the datatype based on unary view in which binary and decimal view are 
derived representations. 
Minus and predecessor are absent in this datatype. 
Successor terms, that is expressions involving zero 
and successor only, serve as normal forms for the datatype $\Nat_{ubd}$.
This DDRS contains the well-known equations $\text{\ref{S1}}-\text{\ref{S4}}$
and the twenty-one equations $\text{\ref{S5}}-\text{\ref{S7}}$, and defines the rewrite rules
that serve the rewriting of binary and decimal notation.

\begin{table}
\hrule
\begin{minipage}[t]{0.4\linewidth}\centering
\begin{Lalign}
\label{S1}
\tag*{[S1]}
x+0 &=x\\
\label{S2}
\tag*{[S2]}
x + S(y) &= S(x+y)\\[2mm]
\label{S3}
\tag*{[S3]}
x \cdot 0 &=0\\
\label{S4}
\tag*{[S4]}
x \cdot S(y) &=(x \cdot y) + x
\end{Lalign}
\end{minipage}\vspace{4mm}
\hfill
\begin{minipage}[t]{0.4\linewidth}\centering
\begin{Lalign}
\label{S5}
\tag*{[S5.$i\hspace{.4mm}]_{i=0}^8$}
i^{\prime} &= S(i)\\
\label{S6}
\tag*{[S6.$i\,]_{i=0}^1$}
x \apb i  &= (x\cdot S(1)) + i\\
\label{S7}
\tag*{[S7.$i\hspace{.4mm}]_{i=0}^9$}
x \apd i &= (x \cdot S(9)) + i
\end{Lalign}
\end{minipage}
\hrule
\caption{A DDRS for~$\Nat_{ubd}$, natural numbers in unary view}
\label{tab:natunview}
\end{table}

In Table~\ref{tab:intunview} a DDRS is provided for the 
$\Int_{ubd}$ of integer numbers with successor, predecessor, addition, and multiplication, 
which are defined by equations~$\text{\ref{u1}}-\text{\ref{u14}}$.
We notice that we do not need equations for rewriting 
\((-x)\cdot y\)
because multiplication is defined by
recursion on its right-argument, and that is why equation~\ref{u11} is sufficient,
and why addition is defined by recursion on
both its arguments and also requires~\ref{u9} and~\ref{u10}.
Like before, the twenty-one equations
$\text{\ref{u15}}-\text{\ref{u17}}$ serve the rewriting of binary and decimal notation.

\begin{table}
\hrule
\begin{minipage}[t]{0.4\linewidth}\centering
\begin{Lalign}
\label{u1}
\tag*{[u1]}
x+0 &=x\\
\label{u2}
\tag*{[u2]}
x + S(y) &= S(x+y)\\[2mm]
\label{u3}
\tag*{[u3]}
x \cdot 0 &=0\\
\label{u4}
\tag*{[u4]}
x \cdot S(y) &=(x \cdot y) + x
\end{Lalign}
\end{minipage}
\hfill
\begin{minipage}[t]{0.49\linewidth}\centering
\begin{Lalign}
\label{u5}
\tag*{[u5]}
-0 &= 0\\
\label{u6}
\tag*{[u6]}
S(-(S(x))) &= -x\\
\label{u7}
\tag*{[u7]}
-(-x) &= x\\[2mm]
\label{u8}
\tag*{[u8]}
0 + x &= x\\
\label{u9}
\tag*{[u9]}
S(x) + y &= S(x+y)\\
\label{u10}
\tag*{[u10]}
(-x) + (-y) &= -(x+y)
\\[2mm]
\label{u11}
\tag*{[u11]}
x \cdot (-y) &= -(x \cdot y)
\\[2mm]
\label{u12}
\tag*{[u12]}
P(0) &= -S(0)
\\
\label{u13}
\tag*{[u13]}
P(S(x)) &= x\\
\label{u14}
\tag*{[u14]}
P(-x) &= -S(x)\\[2mm]
\label{u15}
\tag*{[u15.$i\hspace{.4mm}]_{i=0}^8$}
i^{\prime} &= S(i)\\
\label{u16}
\tag*{[u16.$i\,]_{i=0}^1$}
x \apb i  &= (x\cdot S(1)) + i\\
\label{u17}
\tag*{[u17.$i\hspace{.4mm}]_{i=0}^9$}
x \apd i &= (x \cdot S(9)) + i
\end{Lalign}
\end{minipage}\vspace{4mm}
\hrule
\caption{A DDRS for~$\Int_{ubd}$, integer numbers in unary view}
\label{tab:intunview}
\end{table}

In Table~\ref{tab:EqsForInt} one finds a listing of equations that are true in the 
datatype $\Int_{ubd}$ that
is specified by the DDRS in Table~\ref{tab:intunview}.
This ensures that these equations are semantic 
consequences of the equations for commutative rings.
We give a detailed proof of the ground-completeness of this DDRS for $\Int_{ubd}$
in Appendix~\ref{app:A.1},
which also implies ground-completeness of the DDRS for $\Nat_{ubd}$ defined in 
Table~\ref{tab:natunview}.

\begin{table}
\centering
\hrule
\begin{Ralign}
\label{e1}
x + (y + z) &= (x+y) + z\\
\label{e2}
x+y &= y + x\\
\label{e3}
x+0 &= x\\
\label{e4}
x + (-x) &= 0\\
\label{e5}
(x \cdot y) \cdot z &= x \cdot (y \cdot z) \\
\label{e6}
x \cdot y &= y \cdot x\\
\label{e7}
1\cdot x&=x\\
\label{e8}
x \cdot (y + z) & = (x \cdot y) + (x \cdot z)\\[2mm]
S(x)&=x+1\\
P(x)&=x+(-1)\\[2mm]
x \apb i  &= (x + x) + i&&\text{for $i\in\{0,1\}$}\\[2mm]
x\apd i&=(\underline{10}\cdot x)+\underline i&&\text{for $i\in\{0,1,2,3,4,5,6,7,8,9\}$,}\\
\nonumber
&&&\text{$\underline 0=0$,~~$\underline{i'}=\underline i+1$,~~$\underline{10}=\underline 9+1$}
\end{Ralign}
\hrule
\caption{Equations valid in $\Int_{ubd}$, where $\eqref{e1} - \eqref{e8}$
axiomatize commutative rings}
\label{tab:EqsForInt}
\end{table}

So, binary and decimal notation are defined by expanding terms into successor terms.
This expansion involves a combinatorial explosion in size and
renders the specification in Tables~\ref{tab:natunview} and~\ref{tab:intunview}
irrelevant as term rewriting systems from which an efficient implementation can be generated.

\subsection{Binary view}
\label{sec:2.2}
In Table~\ref{tab:natbinview} we define a DDRS for a binary 
view of natural numbers that employs the successor
function as an auxiliary function.
Leading zeros except for the zero itself are removed by~\ref{b1}, and
successor terms are rewritten according to $\text{\ref{b2}}-\text{\ref{b5}}$.
This DDRS contains fifteen (parametric) equations (that is, sixteen equations for the 
specification of addition and multiplication,
and eighteen that serve the rewriting from decimal notation to binary notation via successor 
terms\footnote{Note that there is no equation [b14.0] that
is, $1=S(0)$, because 1 is a normal form in binary view.}).
In the binary view natural numbers are identified with normal forms in binary notation.
The specification has a canonical term algebra $\Nat_{bud}$ which is isomorphic to the 
canonical term  algebra $\Nat_{ubd}$ of the specification in
Table~\ref{tab:natunview}. In~\cite{KW16}, \textsc{Kluiving} and \textsc{van Woerkom}
prove that this DDRS is complete.

\begin{table}[t]
\centering
\hrule
\begin{minipage}[t]{0.52\linewidth}\centering
\begin{Lalign}
\label{b1}
\tag*{[b1.$i\hspace{.4mm}]_{i=0}^1$}
0\apb i &=i\\[2mm]
\label{b2}
\tag*{[b2]}
S(0) &= 1\\
\label{b3}
\tag*{[b3]}
S(1) &= 1\apbz\\
\label{b4}
\tag*{[b4]}
S(x\apbz) &=x\apbo\\
\label{b5}
\tag*{[b5]}
S(x\apbo) &=S(x)\apbz\\[2mm]
\label{b6}
\tag*{[b6]}
x+0 &=x\\
\label{b7}
\tag*{[b7]}
0 + x &=x\\
\label{b8}
\tag*{[b8]}
x+1 &=S(x)\\
\label{b9}
\tag*{[b9]}
1+x &=S(x)\\
\label{b10}
\tag*{[b10.$i.j\hspace{.4mm}]_{i,j=0}^1$}
(x\apb i) + (y\apb j) &= S^j((x+y)\apb i)
\end{Lalign}
\end{minipage}\vspace{4mm}
\hfill
\begin{minipage}[t]{0.45\linewidth}\centering
\begin{Lalign}
\label{b11}
\tag*{[b11]}
x \cdot 0 &=0\\
\label{b12}
\tag*{[b12]}
x \cdot 1 &=x\\
\label{b13}
\tag*{[b13.$i\hspace{.4mm}]_{i=0}^1$}
x\cdot(y\apb i) &=((x\cdot y)\apbz )+(x\cdot i)\\[2mm]
\label{b14}
\tag*{[b14.$i\hspace{.4mm}]_{i=1}^8$}
i' &= S(i)\\
\label{b15}
\tag*{[b15.$i\hspace{.4mm}]_{i=0}^9$}
x\apd  i &= (x \cdot  S(9)) + i
\end{Lalign}
\end{minipage}
\hrule
\caption{A DDRS for~$\Nat_{bud}$, natural numbers in binary view}
\label{tab:natbinview}
\end{table}

In Table~\ref{tab:binview} minus and predecessor are introduced and the transition from a signature
for natural numbers to a signature for integers is made; the rules in this table extend those
of Table~\ref{tab:natbinview} and define the canonical term algebra $\Int_{bud}$
that is isomorphic to the canonical term algebra $\Int_{ubd}$ of the specification in 
Table~\ref{tab:intunview}.
The DDRS thus defined contains thirty-three (parametric) equations (thus, 34+24 eq's in total).
We attempt to provide some intuition for equations~\ref{b26} and \ref{b27}: 
\[(-x)\apb i\]
should be equal to 
$(-x\apb 0) + i,$
so 
$(-x)\apbz=-( x\apbz)$, and
$(-x) \apbu$ is determined by
\[-(P( x\apbz))\stackrel{\text{\footnotesize\ref{b20}}}=-(P(x) \apbu).\]
Equations~\ref{b24} and \ref{b25} can be explained in a similar way:
\begin{align*}
S(-( x\apbz))&
\quad\text{should be equal to} \quad
-(P(x\apbz))=-(P(x)\apbu),\\
S(-( x\apbu))&
\quad\text{should be equal to} \quad
-(P(x\apbu))=-(x\apbz).
\end{align*}
Normal forms for $\Int_{bud}$
are $0$, $1$, all applications of $\_\apbz$ and $\_\apbu$
to a nonzero normal form, and all minus instances $-t$ of each 
such normal form $t$ that differs from 0.

\begin{table}
\centering
\hrule
\begin{minipage}[t]{0.56\linewidth}\centering
\begin{Lalign}
\tag*{[b1.$i\hspace{.4mm}]_{i=0}^1$}
0\apb i &=i\\[2mm]
\tag*{[b2]}
S(0) &= 1\\
\tag*{[b3]}
S(1) &= 1\apbz\\
\tag*{[b4]}
S(x\apbz) &=x\apbo\\
\tag*{[b5]}
S(x\apbo) &=S(x)\apbz\\[2mm]
\tag*{[b6]}
x+0 &=x\\
\tag*{[b7]}
0 + x &=x\\
\tag*{[b8]}
x+1 &=S(x)\\
\tag*{[b9]}
1+x &=S(x)\\
\tag*{[b10.$i.j\hspace{.4mm}]_{i,j=0}^1$}
(x\apb i) + (y\apb j) &= S^j((x+y)\apb i)
\\[2mm]
\tag*{[b11]}
x \cdot 0 &=0\\
\tag*{[b12]}
x \cdot 1 &=x\\
\tag*{[b13.$i\hspace{.4mm}]_{i=0}^1$}
x\cdot(y\apb i) &=((x\cdot y)\apbz )+(x\cdot i)\\[2mm]
\tag*{[b14.$i\hspace{.4mm}]_{i=1}^8$}
i' &= S(i)\\
\tag*{[b15.$i\hspace{.4mm}]_{i=0}^9$}
x\apd  i &= (x \cdot  S(9)) + i
\end{Lalign}
\end{minipage}
\begin{minipage}[t]{0.43\linewidth}\centering
\begin{Lalign}
\label{b16}
\tag*{[b16]}
-0 &=0\\
\label{b17}
\tag*{[b17]}
-(-x) &= x
\\[3.6mm]
\label{b18}
\tag*{[b18]}
P(0) & =-1\\
\label{b19}
\tag*{[b19]}
P(1) &= 0\\
\label{b20}
\tag*{[b20]}
P(x \apbz) &= P(x)\apbu\\
\label{b21}
\tag*{[b21]}
P(x \apbo) &= x\apbz\\
\label{b22}
\tag*{[b22]}
P(-x) & = -S(x)
\\[3.6mm]
\label{b23}
\tag*{[b23]}
S(-1) &=0\\
\label{b24}
\tag*{[b24]}
S(-( x\apbz)) &= -(P(x) \apbu) \\
\label{b25}
\tag*{[b25]}
S(-(x \apbu)) &= -(x \apbz)
\\[3.6mm]
\label{b26}
\tag*{[b26]}
(-x) \apbz &= -(x \apbz)\\
\label{b27}
\tag*{[b27]}
(-x) \apbu &= -(P(x) \apbu)
\\[3.6mm]
\label{b28}
\tag*{[b28]}
x + (-1) &= P(x)\\
\label{b29}
\tag*{[b29]}
(-1) + x &= P(x)
\end{Lalign}
\end{minipage}
\center{\begin{minipage}[t]{0.64\linewidth}\centering
\begin{Lalign}
\label{b30}
\tag*{[b30.$i.j\hspace{.4mm}]_{i,j=0}^1$}
(x\apb i) + (-(y\apb j)) &= P^j((x+(-y))\apb i)\\
\label{b31}
\tag*{[b31.$i.j\hspace{.4mm}]_{i,j=0}^1$}
(-(y\apb j)) + (x\apb i) &= P^j((x+(-y))\apb i)\\
\label{b32}
\tag*{[b32]}
(-x) + (-y) &= -(x+y)\\[2mm]
\label{b33}
\tag*{[b33]}
x \cdot (-y) &= -(x \cdot y)
\end{Lalign}
\end{minipage}}
\vspace{4mm}
\hrule
\caption{A DDRS for~$\Int_{bud}$ that specifies 
integer numbers in binary view}
\label{tab:binview}
\end{table}

We note that the equations in Table~\ref{tab:binview} are semantic 
consequences of the axioms for commutative rings 
(equations $\eqref{e1} - \eqref{e8}$ in Table~\ref{tab:EqsForInt}).
This DDRS is proven strongly terminating in~\cite{KW16}.
However, its non-confluence is also proven in~\cite{KW16}, using the following rewrite
steps: 
\begin{Ralign}
\label{c-exA}
\tag{13}
\begin{picture}(40,58)(30,-6)
\put(20,40){$P(-(-x))$}
\put(34,30){\vector(-3,-2){34}}
\put(-22,-5){$P(x)$}
\put(48,30){\vector(3,-2){34}}
\put(80,-5){$-S(-x)$}
\put(-8,19){\text{\small\ref{b17}}}
\put(70,19){\text{\small\ref{b22}}}
\end{picture}
\end{Ralign}
In Appendix~\ref{app:A.2} we prove that this DDRS for $\Int_{bud}$ is ground-confluent,
and thus ground-complete. 

\subsection{Decimal view}
\label{sec:2.3}
In Table~\ref{tab:decview} we define a DDRS for a decimal 
view of natural numbers that defines the canonical term algebra 
$\Nat_{dub}$, the datatype in which unary and binary view are 
derived representations. This DDRS consists of fourteen (parametric) equations 
(172 eq's in total). 
The datatype $\Nat_{dub}$  is isomorphic to the canonical term  algebra $\Nat_{ubd}$ of 
the specification in Table~\ref{tab:natunview}. 
Leading zeros except for the zero itself are removed by \ref{d1}, and
successor terms are rewritten according to $\text{\ref{d2}}-\text{\ref{d5}}$.
Rewriting from binary notation is part of this DDRS, and the last equation scheme~\ref{d14}
serves that purpose. In~\cite{KW16}, this DDRS for $\Nat_{dub}$ is proven complete.

\begin{table}
\centering
\hrule
\begin{minipage}[t]{0.51\linewidth}\centering
\begin{Lalign}
\label{d1}
\tag*{[d1.$i\hspace{.4mm}]_{i=0}^9$}
0 \apd i &= i\\[2mm]
\label{d2}
\tag*{[d2.$i\hspace{.4mm}]_{i=0}^8$}
S(i) &=i^{\prime}\\
\label{d3}
\tag*{[d3]}
S(9) &= 1 \apd 0\\
\label{d4}
\tag*{[d4.$i\hspace{.4mm}]_{i=0}^8$}
S(x \apd i) &= x \apd i^{\prime}\\
\label{d5}
\tag*{[d5]}
S(x \apd 9) &= S(x) \apd 0\\[2mm]
\label{d6}
\tag*{[d6]}
x + 0 &=x\\
\label{d7}
\tag*{[d7]}
0 + x &=x\\
\label{d8}
\tag*{[d8.$i\hspace{.4mm}]_{i=1}^9$}
x+i &= S^i(x)\\
\label{d9}
\tag*{[d9.$i\hspace{.4mm}]_{i=1}^9$}
i + x &= S^i(x)\\
\label{d10}
\tag*{[d10.$i.j\hspace{.4mm}]_{i,j=0}^9$}
(x \apd i)+(y \apd j) &= S^j((x +y) \apd i)
\end{Lalign}
\end{minipage}\vspace{4mm}
\hfill
\begin{minipage}[t]{0.48\linewidth}\centering
\begin{Lalign}
\label{d11}
\tag*{[d11]}
x \cdot 0 &=0\\
\label{d12}
\tag*{[d12.$i\hspace{.4mm}]_{i=0}^8$}
x \cdot i^{\prime}  & = (x \cdot i) + x\\
\label{d13}
\tag*{[d13.$i\hspace{.4mm}]_{i=0}^9$}
x \cdot (y \apd i) &= ((x \cdot y) \apd 0) + (x \cdot i)\\[2mm]
\label{d14}
\tag*{[d14.$i\hspace{.4mm}]_{i=0}^1$}
x\apb  i &= (x +x) + i
\end{Lalign}
\end{minipage}
\hrule
\caption{A DDRS for~$\Nat_{dub}$, natural numbers in decimal view}
\label{tab:decview}
\end{table}

Before we extend the DDRS in Table~\ref{tab:decview} to the integers, we define 
in Table~\ref{fig:subdig} a variant of successor notation for digits
that we call ``10 minus subtraction'' with notation $i^\star$, and that for decimal digits 
$i\in\{1,...,9\}$ characterizes the equation
\[i^\star=10-i.\]
In Table~\ref{tab:indecview}, minus and predecessor are added and the transition 
to integers is made.
In rule scheme~\ref{d26} we employ the notation $i^\star$.
The DDRS thus defined is named $\Int_{dub}$ and is isomorphic to the canonical term algebra 
$\Int_{ubd}$ of the specification in Table~\ref{tab:intunview}; it contains thirty-two 
(parametric) equations (so, $172 + 272$ eq's in total).

\begin{table}[b]
\centering
\hrule
\begin{align*}
1^{\star} &\equiv 9&4^{\star} &\equiv 6&7^{\star} &\equiv 3\\
2^{\star} &\equiv 8&5^{\star} &\equiv 5&8^{\star} &\equiv 2\\
3^{\star} &\equiv 7&6^{\star} &\equiv 4&9^{\star} &\equiv 1
\end{align*}
\hrule
\caption{``10 minus'' subtraction notation for decimal digits}
\label{fig:subdig}
\end{table}

The (twenty) equations captured by $\text{\ref{d23}}-\text{\ref{d26} }$ can be explained
in a similar fashion as was done in the previous section for $\text{\ref{b24}}-\text{\ref{b27}}$: 
for example, 
\[(-5)\apd3\]
should be equal to $-(5\apd0)+3=-(4\apd 7)$, and this follows immediately from the appropriate
equation in~\ref{d26}.

The equations of the DDRS specified by~Tables~\ref{tab:decview}
and \ref{tab:indecview} are semantic 
consequences of the equations for commutative rings 
(equations $\eqref{e1} - \eqref{e8}$ in Table~\ref{tab:EqsForInt}).
In~\cite{KW16}, this DDRS for $\Int_{dub}$ is proven strongly terminating, and 
non-confluent by essentially the same counter-example as was used for the DDRS for 
$\Int_{bud}$ (see~\eqref{c-exA}):
\[
\begin{picture}(40,68)(30,-12)
\put(20,40){$P(-(-x))$}
\put(34,30){\vector(-3,-2){34}}
\put(-22,-5){$P(x)$}
\put(48,30){\vector(3,-2){34}}
\put(80,-5){$-S(-x)$}
\put(-8,19){\text{\small\ref{d16}}}
\put(70,19){\text{\small\ref{d21}}}
\end{picture}
\]
In Appendix~\ref{app:A.3} we prove that this DDRS for $\Int_{dub}$ is ground-confluent,
and thus ground-complete.

\begin{table}[h]
\hrule
\begin{minipage}[t]{0.56\linewidth}\centering
\begin{Lalign}
\tag*{[d1.$i\hspace{.4mm}]_{i=0}^9$}
0 \apd i &= i\\[2mm]
\tag*{[d2.$i\hspace{.4mm}]_{i=0}^8$}
S(i) &=i^{\prime}\\
\tag*{[d3]}
S(9) &= 1 \apd 0\\
\tag*{[d4.$i\hspace{.4mm}]_{i=0}^8$}
S(x \apd i) &= x \apd i^{\prime}\\
\tag*{[d5]}
S(x \apd 9) &= S(x) \apd 0\\[2mm]
\tag*{[d6]}
x + 0 &=x\\
\tag*{[d7]}
0 + x &=x\\
\tag*{[d8.$i\hspace{.4mm}]_{i=1}^9$}
x+i &= S^i(x)\\
\tag*{[d9.$i\hspace{.4mm}]_{i=1}^9$}
i + x &= S^i(x)\\
\tag*{[d10.$i.j\hspace{.4mm}]_{i,j=0}^9$}
(x \apd i)+(y \apd j) &= S^j((x +y) \apd i)
\\[2mm]
\tag*{[d11]}
x \cdot 0 &=0\\
\tag*{[d12.$i\hspace{.4mm}]_{i=0}^8$}
x \cdot i^{\prime}  & = (x \cdot i) + x\\
\tag*{[d13.$i\hspace{.4mm}]_{i=0}^9$}
x \cdot (y \apd i) &= ((x \cdot y) \apd 0) + (x \cdot i)\\[2mm]
\tag*{[d14.$i\hspace{.4mm}]_{i=0}^1$}
x\apb  i &= (x +x) + i
\end{Lalign}
\end{minipage}
\begin{minipage}[t]{0.43\linewidth}\centering
\begin{Lalign}
\label{d15}
\tag*{[d15]}
-0 &=0\\
\label{d16}
\tag*{[d16]}
-(-x) &= x
\\[2.2mm]
\label{d17}
\tag*{[d17]}
P(0)&=-1\\
\label{d18}
\tag*{[d18.$i\hspace{.4mm}]_{i=0}^8$}
P(i^\prime)  &=  i\\
\label{d19}
\tag*{[d19]}
P(x \apd 0) &= P(x) \apd 9\\
\label{d20}
\tag*{[d20.$i\hspace{.4mm}]_{i=0}^8$}
P(x \apd i^\prime) &= x \apd i\\
\label{d21}
\tag*{[d21]}
P(-x) &= -S(x)
\\[2.2mm]
\label{d22}
\tag*{[d22.$i\hspace{.4mm}]_{i=0}^8$}
S(-i^\prime)  &= - i\\
\label{d23}
\tag*{[d23]}
S(-(x \apd 0)) &= -(P(x) \apd 9)\\
\label{d24}
\tag*{[d24.$i\hspace{.4mm}]_{i=0}^8$}
S(-(x \apd i^\prime)) &= -(x \apd i)
\\[2.2mm]
\label{d25}
\tag*{[d25]}
(-x) \apd 0 &= -(x \apd 0)\\[0mm]
\label{d26}
\tag*{[d26.$i\hspace{.4mm}]_{i=1}^9$}
(-x) \apd i &= -(P(x) \apd i^{\star})
\\[2.2mm]
\label{d27}
\tag*{[d27.$i\hspace{.4mm}]_{i=1}^9$}
x + (- i) &= P^i(x)\\
\label{d28}
\tag*{[d28.$i\hspace{.4mm}]_{i=1}^9$}
(- i) + x &= P^i(x)
\end{Lalign}
\end{minipage}
\center{
\begin{minipage}[t]{0.58\linewidth}
\begin{Lalign}
\label{d29}
\tag*{[d29.$i.j\hspace{.4mm}]_{i,j=0}^9$}
(x\apd i) + (-(y\apd j)) &= P^j((x+(-y))\apd i )\\
\label{d30}
\tag*{[d30.$i.j\hspace{.4mm}]_{i,j=0}^9$}
(-(y\apd j)) + (x\apd i) &= P^j((x+(-y))\apd i )\\
\label{d31}
\tag*{[d31]}
(-x) + (-y) &= -(x+y)\\[2mm]
\label{d32}
\tag*{[d32]}
x \cdot (-y) &= -(x \cdot y)
\end{Lalign}
\end{minipage}}
\vspace{4mm}
\hrule
\caption{A DDRS for~$\Int_{dub}$ that specifies integers in decimal view, employing $i^\star$ from 
Table~\ref{fig:subdig}}
\label{tab:indecview}
\end{table}

\newpage

\section{Alternative views with digit tree constructors}
\label{sec:3}

Having defined DDRSes that employ (postfix) digit append functions in Section~\ref{sec:2},
we now consider the more general \emph{digit tree constructor} functions. 
For the binary view, this approach is followed by \textsc{Bouma} and 
\textsc{Walters} in~\cite{BW89}; for a view
based on any radix (number base), this approach is further continued in 
\textsc{Walters}~\cite{Wal94} and \textsc{Walters} and 
\textsc{Zantema}~\cite{WZ95}, where the constructor is called \emph{juxtaposition} 
because it goes with the absence of a function symbol in order to be close to ordinary decimal 
and binary notation.  

We extend the signature $\Sigma_\Int$ defined in Section~\ref{sec:sig} with the following three 
functions (infix):
\[\concu,\concb, \concd: \Int\times \Int \to \Int,\]
called ``unary digit tree constructor function'', ``binary digit tree constructor function'',
and ``decimal digit tree constructor function'',
and to be used for unary, binary notation and decimal notation, respectively.
The latter two constructors serve to represent 
positional notation and satisfy the semantic equations 
$\llbracket x\concb y\rrbracket=2\cdot\llbracket x\rrbracket+\llbracket y\rrbracket$
and $\llbracket x\concd y\rrbracket=10\cdot\llbracket x\rrbracket+\llbracket y\rrbracket$.

For integer numbers in decimal view or binary view, normal forms are the relevant digits, 
all applications of the respective constructor with left argument a nonzero normal form and 
right argument a digit, and all minus instances $-t$ of each such nonzero normal form $t$, 
these satisfy $\llbracket-(t)\rrbracket=-(\llbracket t\rrbracket)$.
E.g.,
\[(9\concd7)\concd5\quad \text{and}\quad ((1\concb0)\concb 0)\concb 1\]
represent the decimal number 975
and the binary number $1001$, respectively, and the normal form that represents the additional 
inverse of the latter is $-(((1\concb0)\concb0)\concb1)$.
A minor complication with decimal and binary digit tree constructors is that we now have to
consider rewritings such as
\[2\concd (1\concd 5)=(2+1)\concd 5=3\concd 5 \quad(=35),\]
which perhaps are somewhat non-intuitive. For integers in unary view, thus with 
unary digit tree constructor, this complication is absent (see Section~\ref{subsec:3.1}).

We keep the presentation of the resulting DDRSes (those defining the binary and decimal view 
are based on~\cite{Wal94,WZ95}) 
minimal in the sense that equations for conversion from the one view
to the other are left out. Of course, it is easy to define such equations.
Also, equations for conversion to and from the datatypes defined in Section~\ref{sec:2} are 
omitted, although such equations are also easy to define.
 
\subsection{Unary view with digit tree constructor}
\label{subsec:3.1}
For naturals in this particular unary view, normal forms are 0 and expressions $t\concu 0$ 
with $t$ a normal form (thus, with association of $\concu$ to the left).
Of course, the phenomenon of ``removing leading zeros'' does not exist in this particular unary 
view. The resulting datatype $\Nat_{ut}$ is defined in Table~\ref{tab:natuaview}.

\begin{table}
\centering
\hrule
\begin{minipage}[t]{0.49\linewidth}\centering
\begin{Lalign}
\label{ut1}
\tag*{[ut1]}
x\concu(y\concu z)&=(x\concu y)\concu z\\[2mm]
\label{ut2}
\tag*{[ut2]}
x+0 &=x\\
\label{ut3}
\tag*{[ut3]}
x + (y\concu 0)&= (x+y)\concu 0
\end{Lalign}
\end{minipage}\vspace{4mm}
\hfill
\begin{minipage}[t]{0.49\linewidth}\centering
\begin{Lalign}
\label{ut4}
\tag*{[ut4]}
x \cdot 0 &=0\\
\label{ut5}
\tag*{[ut5]}
x \cdot (y\concu 0)&= (x\cdot y)+x
\end{Lalign}
\end{minipage}
\hrule
\caption{A DDRS for~$\Nat_{ut}$, natural numbers in unary view with unary digit tree constructor}
\label{tab:natuaview}
\end{table}

In the unary view, $\concu$ is an associative operator, as is clear from rule~\ref{ut1} 
(in contrast to digit tree constructors for the binary and decimal case).
Moreover, the commutative variants $t\concu r$ and $r\concu t$ rewrite to the same normal form.
The latter property also follows from the following  semantics for closed terms:
\begin{align*}
\llbracket 0 \rrbracket&=0,\\
\llbracket x\concu y \rrbracket&=\llbracket x \rrbracket + \llbracket y \rrbracket +1,\\
\llbracket x+ y \rrbracket&=\llbracket x \rrbracket + \llbracket y \rrbracket,\\
\llbracket x\cdot y \rrbracket&=\llbracket x \rrbracket \cdot \llbracket y \rrbracket .
\end{align*}
Observe that 
\[x + (y\concu z) = (x+y)\concu z\quad
\text{and}\quad x\cdot (y\concu z) =(x \cdot (y+z)) + x\]
are valid equations in $\Nat_{ut}$.

The extension to integer numbers can be done in a similar fashion as in the previous section,
thus obtaining normal forms of the form $-(t)$ with $t$ a nonzero normal form in $\Nat_{ut}$.
However, also terms of the form $x\concu (-y)$ and variations thereof have to be considered.
We define this extension in Table~\ref{tab:intuaview} below and call the resulting datatype 
$\Int_{ut}$.

Adding the interpretation rule
$\llbracket -x \rrbracket=-\llbracket x  \rrbracket
$ and exploiting the commutativity of $\concu$ in $\llbracket x\concu y \rrbracket$, 
it can be easily checked that $\text{\ref{ut6}}-\text{\ref{ut16}}$ (as equations) are sound. 
In Appendix~\ref{app:B.1} we prove that this DDRS is ground-complete,
which also implies ground-completeness of the DDRS for $\Nat_{ut}$ defined in 
Table~\ref{tab:natuaview}: strong termination is preserved and
all its equations are valid.

\begin{table}[h]
\centering
\hrule
\begin{minipage}[t]{0.41\linewidth}\centering
\begin{Lalign}
\tag*{[ut1]}
x\concu(y\concu z)&=(x\concu y)\concu z\\[2mm]
\tag*{[ut2]}
x+0 &=x\\
\tag*{[ut3]}
x + (y\concu 0)&= (x+y)\concu 0
\\[2mm]
\tag*{[ut4]}
x \cdot 0 &=0\\
\tag*{[ut5]}
x \cdot (y\concu 0)&= (x\cdot y)+x
\end{Lalign}
\end{minipage}
\begin{minipage}[t]{0.58\linewidth}\centering
\begin{Lalign}
\label{ut6}
 \tag*{[ut6]}
-0&=0\\
\label{ut7}
\tag*{[ut7]}
-(-x) &=x\\[2mm]
\label{ut8}
\tag*{[ut8]}
0 \concu (-(x\concu 0)) &= -x
\\
\label{ut9}
\tag*{[ut9]}
(x\concu 0) \concu (-(y\concu 0)) &= x\concu (-y)
\\
\label{ut10}
\tag*{[ut10]}
(-(x\concu 0)) \concu 0 &= -x\\
\label{ut11}
\tag*{[ut11]}
(-(x\concu 0))\concu(y\concu 0)&= (-x)\concu y\\
\label{ut12}
\tag*{[ut12]}
(-(x\concu 0)) \concu (-(y\concu 0)) &= -((x+y)\concu 0)
\\[2mm]
\label{ut13}
\tag*{[ut13]}
0+x&=x
\\
\label{ut14}
\tag*{[ut14]}
(x\concu 0)+(-(y\concu 0))&=x+(-y)\\
\label{ut15}
\tag*{[ut15]}
(-x) + (-y) &=-(x+ y)
\\[2mm]
\label{ut16}
\tag*{[ut16]}
x \cdot (-y) &=-(x\cdot y)
\end{Lalign}
\end{minipage}\vspace{4mm}
\hrule
\caption{A DDRS for~$\Int_{ut}$ that specifies 
integer numbers in unary view with unary digit tree constructor}
\label{tab:intuaview}
\end{table}

\subsection{Binary view with digit tree constructor}
\label{subsec:3.2}
For naturals in binary view with the binary digit tree constructor, the associated 
datatype $\Nat_{bt}$ is defined in Table~\ref{tab:bc}. According to~\cite{WZ95}
(with a reference to~\cite{BW89}),
the rewriting system defined by $\text{\ref{bt1}}-\text{\ref{bt7}}$
is strongly terminating and ground-confluent, and 
thus ground-complete. 

\begin{table}
\centering
\hrule
\begin{minipage}[t]{0.45\linewidth}\centering
\begin{Lalign}
\label{bt1}
\tag*{[bi1]}
0\concb x&=x\\
\label{bt2}
\tag*{[bi2]}
x\concb(y\concb z)&=(x+ y)\concb z
\\[2mm]
\label{bt3}
\tag*{[bi3]}
0+x &=x\\
\label{bt4}
\tag*{[bi4]}
1+0 &=1\\
\label{bt5}
\tag*{[bi5]}
1+1 &=1\concb 0
\\
\label{bt6}
\tag*{[bi6]}
1 + (x\concb y) &= x\concb(1+y)\\
\label{bt7}
\tag*{[bi7]}
(x \concb y)+z &= x\concb(y+z)
\end{Lalign}
\end{minipage}\vspace{3mm}
\hfill
\begin{minipage}[t]{0.45\linewidth}\centering
\begin{Lalign}
\label{bt8}
\tag*{[bi8]}
x \cdot 0 &= 0\\
\label{bt9}
\tag*{[bi9]}
x \cdot 1 &= x\\
\label{bt10}
\tag*{[bi10]}
x \cdot (y\concb z) &= (x\cdot y)\concb (x\cdot z)
\end{Lalign}
\end{minipage}
\hrule
\caption{A DDRS for~$\Nat_{bt}$, natural numbers in binary view with binary digit tree constructor}
\label{tab:bc}
\end{table}

In~\cite{WZ95} a rewriting system for integer arithmetic is provided with next to juxtaposition 
and minus also addition, subtraction and multiplication, and proven ground-confluent and 
terminating with respect to any radix (number base). 
In Table~\ref{natbview} we present a variant
of this rewriting system without subtraction for the binary digit tree constructor, and 
define the datatype $\Int_{bt}$. 

\bigskip

\begin{table}[h]
\hrule
\begin{minipage}[t]{0.4\linewidth}\centering
\begin{Lalign}
\label{bi1}
\tag*{[bt1]}
0\concb x&=x\\
\label{bi2}
\tag*{[bt2]}
x\concb(y\concb z)&=(x+ y)\concb z\\[2mm]
\label{bi3}
\tag*{[bt3]}
0+x &=x\\
\label{bi4}
\tag*{[bt4]}
x+0 &=x\\
\label{bi5}
\tag*{[bt5]}
1+1 &=1\concb 0\\
\label{bi6}
\tag*{[bt6]}
x + (y\concb z) &= y\concb(x+z)\\
\label{bi7}
\tag*{[bt7]}
(x \concb y)+z &= x\concb(y+z)
\\[2mm]
\label{bi8}
\tag*{[bt8]}
x \cdot 0 &=0\\
\label{bi9}
\tag*{[bt9]}
0\cdot x&=0\\
\label{bi10}
\tag*{[bt10]}
1\cdot 1&=1\\
\label{bi11}
\tag*{[bt11]}
x \cdot (y\concb z) &=(x \cdot y)\concb(x\cdot z)\\
\label{bi12}
\tag*{[bt12]}
(x \concb y)\cdot z &=(x \cdot z)\concb(y\cdot z)
\end{Lalign}
\end{minipage}\vspace{4mm}
\hfill
\begin{minipage}[t]{0.54\linewidth}\centering
\begin{Lalign}
\label{bi13}
\tag*{[bt13]}
-0&=0\\
\label{bi14}
\tag*{[bt14]}
-(-x)&=x\\[2mm]
\label{bi15}
\tag*{[bt15]}
1\concb (-1)&=1\\
\label{bi16}
\tag*{[bt16]}
(x\concb 0)\concb (-1)&=(x\concb(-1))\concb1\\
\label{bi17}
\tag*{[bt17]}
(x\concb 1)\concb (-1)&=(x\concb0)\concb 1\\
\label{bi18}
\tag*{[bt18]}
x\concb (-(y\concb z))&=-((y+(-x))\concb z)\\
\label{bi19}
\tag*{[bt19]}
(-x)\concb y&=-(x\concb(-y))\\[2mm]
\label{bi20}
\tag*{[bt20]}
1+(-1)&=0\\
\label{bi21}
\tag*{[bt21]}
(-1)+1&=0\\
\label{bi22}
\tag*{[bt22]}
(-1)+(-1)&=-(1\concb 0)\\
\label{bi23}
\tag*{[bt23]}
x+(-(y\concb z))&=-(y\concb (z+(-x)))\\
\label{bi24}
\tag*{[bt24]}
(-(x\concb y))+z&=-(x\concb (y+(-z)))\\[2mm]
\label{bi25}
\tag*{[bt25]}
x\cdot (-y)&=-(x\cdot y)\\
\label{bi26}
\tag*{[bt26]}
(-x)\cdot y&=-(x\cdot y)
\end{Lalign}
\end{minipage}
\hrule
\caption{A DDRS for~$\Int_{bt}$, integer numbers in binary view with binary digit tree constructor}
\label{natbview}
\end{table}

\newpage

In~\cite{KW16} it is proven that the associated term rewriting system is strongly terminating.
Confluence is disproven in~\cite{KW16} by the following counter-example:
\begin{Ralign}
\label{c-ex}
\tag{14}
\begin{picture}(40,110)(20,-56)
\put(10,40){$x\concb(y\concb (z\concb w))$}
\put(34,30){\vector(-3,-2){34}}
\put(-62,-5){$(x+y)\concb(z\concb w)$}
\put(48,30){\vector(3,-2){34}}
\put(78,-5){$x\concb((y+z)\concb w)$}
\put(-8,19){\text{\small\ref{bi2}}}
\put(70,19){\text{\small\ref{bi2}}}
\put(-62,-50){$((x+y)+z)\concb w$}
\put(78,-50){$(x+(y+z))\concb w$}
\put(-18,-14){\vector(0,-1){24}}
\put(100,-14){\vector(0,-1){24}}
\put(-38,-28){\text{\small\ref{bi2}}}
\put(102,-28){\text{\small\ref{bi2}}}
\end{picture}
\end{Ralign}
However, ground-confluence for this DDRS is proven in 
Appendix~\ref{app:B.2}, by which it is ground-complete.
As a consequence, equations $\text{\ref{bi1}}-\text{\ref{bi12}}$ define an alternative
DDRS for $\Nat_{bt}$ that is also ground-complete:
strong termination is preserved and
all equations are valid.

\newpage

\subsection{Decimal view with digit tree constructor}
\label{sec:4.3}
For naturals in decimal view with the decimal digit tree constructor,
we make use of
successor terms, in order to avoid (non-parametric) equations such as 
\[\begin{aligned}
&1+1=2,~\dots,&9+8=1\concd 7,&&9+9 = 1\concd 8,\\ 
&\dots,\\
&1\cdot 1=1,~~\dots,&8\cdot 9=7\concd 2,&&~9\cdot 9=8\concd1.~\end{aligned}\]
The associated datatype $\Nat_{dt}$ is defined in Table~\ref{tab:natdecview}.
Following \textsc{van der Kamp}~\cite{Kamp16}, we use in equations~\ref{dt10}
for  $i\in\{1,2,...,9\}$ the notation
\[\textstyle\sum^i x\]
for $i-1$  repeated applications of $+$ with association to the right, thus
\[\textstyle\sum^1 x=x \quad\text{and for $i=1,...,8$}, 
\quad\textstyle\sum^{i+1}x=x+\sum^i x.\]

\begin{table}
\centering
\hrule
\begin{minipage}[t]{0.4\linewidth}\centering
\begin{Lalign}
\label{dt1}
\tag*{[dt1]}
0\concd x&=x\\
\label{dt2}
\tag*{[dt2]}
x\concd (y\concd z)&=(x+y)\concd z\\[2mm]
\label{dt3}
\tag*{[dt3.$i\hspace{.4mm}]_{i=0}^8$}
S(i)&=i'\\
\label{dt4}
\tag*{[dt4]}
S(9)&=1\concd 0\\
\label{dt5}
\tag*{[dt5.$i\hspace{.4mm}]_{i=0}^8$}
S(x\concd i)&=x\concd i'\\
\label{dt6}
\tag*{[dt6]}
S(x\concd 9)&=S(x)\concd 0
\end{Lalign}
\end{minipage}\vspace{4mm}
\hfill
\begin{minipage}[t]{0.5\linewidth}\centering
\begin{Lalign}
\label{dt7}
\tag*{[dt7.$i\hspace{.4mm}]_{i=0}^9$}
x+i&=S^i(x)\\
\label{dt8}
\tag*{[dt8.$i\hspace{.4mm}]_{i=0}^9$}
x+(y\concd i)&=S^i(y\concd x)\\[2mm]
\label{dt9}
\tag*{[dt9]}
x\cdot 0 &=0\\
\label{dt10}
\tag*{[dt10.$i\hspace{.4mm}]_{i=1}^9$}
x\cdot i&=\textstyle\sum^i x\\
\label{dt11}
\tag*{[dt11.$i\hspace{.4mm}]_{i=0}^9$}
x\cdot (y\concd i)&=((x\cdot y)\concd 0)+(x\cdot i)
\end{Lalign}
\end{minipage}
\hrule
\caption{A DDRS for~$\Nat_{dt}$, natural numbers with decimal digit tree constructor in decimal view
(using the notation $i^\prime$ from Table~\ref{fig:enumdig})}
\label{tab:natdecview}
\end{table}

The extension to integers is given by the equations in Table~\ref{tab:intdecview} that
define the datatype $\Int_{dt}$.
In contrast to the approaches in~\cite{Wal94,WZ95} with juxtaposition, we now make use of
both successor terms and 
predecessor terms, and the DDRS presented here is composed from rewrite  
rules for successor and predecessor, rewrite rules defined in~\cite{Wal94,WZ95}, 
and combinations thereof. 
For a smooth, parametric representation we also use the predecessor notation $i''$ for  
digits larger than $0$ defined in Table~\ref{tab:predig}.
In~\cite{Kamp16} it is shown that the associated rewriting system for $\Int_{dt}$ is strongly 
terminating and ground-confluent, and thus ground-complete.
This implies that the DDRS in Table~\ref{tab:natdecview} for $\Nat_{dt}$ is also ground-complete: 
strong termination is preserved and
all its equations are valid.
Finally we note that both these DDRSes are not confluent 
(cf.~counter-example~\eqref{c-ex}).
\begin{table}[b]
\centering
\hrule
\begin{align*}
1'' &\equiv 0&4'' &\equiv 3&7'' &\equiv 6\\
2'' &\equiv 1&5'' &\equiv 4&8'' &\equiv 7\\
3'' &\equiv 2&6'' &\equiv 5&9'' &\equiv 8
\end{align*}
\hrule
\caption{Predecessor notation for decimal digits}
\label{tab:predig}
\end{table}

\begin{table}[t]
\hrule
\begin{minipage}[t]{0.47\linewidth}\centering
\begin{Lalign}
\tag*{\ref{dt1}}
0\concd x&=x\\
\tag*{\ref{dt2}}
x\concd (y\concd z)&=(x+y)\concd z\\[2mm]
\tag*{\ref{dt3}}
S(i)&=i'\\
\tag*{\ref{dt4}}
S(9)&=1\concd 0\\
\tag*{\ref{dt5}}
S(x\concd i)&=x\concd i'\\
\tag*{\ref{dt6}}
S(x\concd 9)&=S(x)\concd 0\\[2mm]
\tag*{\ref{dt7}}
x+i&=S^i(x)\\
\tag*{\ref{dt8}}
x+(y\concd i)&=S^i(y\concd x)\\[2mm]
\tag*{[dt9]}
x\cdot 0 &=0\\
\tag*{[dt10.$i\hspace{.4mm}]_{i=1}^9$}
x\cdot i&=\textstyle\sum^i x\\
\tag*{[dt11.$i\hspace{.4mm}]_{i=0}^9$}
x\cdot (y\concd i)&=((x\cdot y)\concd 0)+(x\cdot i)
\end{Lalign}
\end{minipage}\vspace{4mm}
\begin{minipage}[t]{0.52\linewidth}\centering
\begin{Lalign}
\label{dt12}
\tag*{[dt12]}
-0&=0\\
\label{dt13}
\tag*{[dt13]}
-(-x)&=x
\\[2.6mm]
\label{dt14}
\tag*{[dt14]}
P(0)&=-1\\
\label{dt15}
\tag*{[dt15.$i\hspace{.4mm}]_{i=0}^8$}
P(i^\prime)  &=  i\\
\label{dt16}
\tag*{[dt16]}
P(x \concd 0) &= P(x) \concd 9\\
\label{dt17}
\tag*{[dt17.$i\hspace{.4mm}]_{i=0}^8$}
P(x \concd i^\prime) &= x \concd i
\\
\label{dt18}
\tag*{[dt18]}
P(-x) &= -S(x)
\\[2.6mm]
\label{dt19}
\tag*{[dt19.$i\hspace{.4mm}]_{i=0}^8$}
S(-i^\prime)  &= - i\\
\label{dt20}
\tag*{[dt20]}
S(-(x \concd 0)) &= -(P(x) \concd 9)\\
\label{dt21}
\tag*{[dt21.$i\hspace{.4mm}]_{i=0}^8$}
S(-(x \concd i^\prime)) &= -(x \concd i)
\\[2.6mm]
\label{dt22}
\tag*{[dt22]}
(-x)\concd y&=-(x\concd(-y))\\[2mm]
\label{dt23}
\tag*{[dt23.$i.j\hspace{.4mm}]_{i,j=1}^9$}
i\concd (-j)&=i''\concd j^\star\\
\label{dt24}
\tag*{[dt24.$i.j\hspace{.4mm}]_{i,j=1}^9$}
(x\concd i)\concd (-j)&=(x\concd i'')\concd j^\star\\
\label{dt25}
\tag*{[dt25]}
x\concd (-(y\concd z))&=-((y+(-x))\concd z)\\[2mm]
\label{dt26}
\tag*{[dt26.$i\hspace{.4mm}]_{i=1}^9$}
x+(-i)&=P^i(x)\\
\label{dt27}
\tag*{[dt27]}
x+(-(y\concd z))&=-(y\concd (z+(-x)))
\\[2mm]
\label{dt28}
\tag*{[dt28]}
x\cdot (-y)&=-(x\cdot y)
\end{Lalign}
\end{minipage}
\hrule
\caption{A DDRS for~$\Int_{dt}$, integer numbers with decimal digit tree constructor in 
decimal view
(using $i^\prime$ from Table~\ref{fig:enumdig}, $i^\star$ from Table~\ref{fig:subdig},
and $i''$ from Table~\ref{tab:predig})}
\label{tab:intdecview}
\end{table}

\section{Concluding remarks}
\label{sec:4}
This paper is about the design (by means of trial and error)  
of \emph{datatype defining rewrite systems} (DDRSes) rather than about the precise analysis
of the various rewriting systems per se.  
What matters in addition to readability and conciseness of each DDRS is at this stage 
a proof | or at least a reasonable confidence |
that each of these rewriting systems is strongly terminating and 
ground-confluent (and thus ground-complete), and furthermore that the
(intended) normal forms are natural and convincing, while the rewriting
systems are comprehensible. 

When specifying a datatype of integers as an extension of the naturals, the unary view
leads to satisfactory results, but with high inefficiency. 
For the binary view and the decimal view based on the unary append functions and discussed
in Section~\ref{sec:2}, such 
extensions are provided, but the resulting rewriting systems are at first sight 
significantly less concise and comprehensible. 
Recently, strong termination has been proven by \textsc{Kluiving} and 
\textsc{van Woerkom}~\cite{KW16} with help of the AProVE tool~\cite{aprove}, and 
ground-confluence is proven in this paper. 
Some further remarks:

\begin{enumerate}
\item The three DDRSes (datatype defining rewrite systems) for integers given in 
Section~\ref{sec:2} each produce an extension datatype for a datatype for the
natural numbers. 
An initial algebra specification of the datatype of integers is obtained from 
any of the DDRSes given in~\cite{Bergstra2014n} by 
\begin{itemize}
\item taking the reduct to the signature involving unary, binary, and decimal notation only,
\item removing rewrite rules involving operators for hexadecimal notation,
\item expanding the signature with a unary additive inverse and a unary predecessor function,
\item adding rewrite rules (in equational form) that allow for the unique normalization of 
closed terms involving the minus sign,
\end{itemize}
while making sure that these rewrite rules (viewed as equations) are semantic 
consequences of the equations for commutative rings.

\item Syntax for hexadecimal notation has been omitted because that usually plays no role 
when dealing with integers. It is an elementary  
exercise to incorporate hexadecimal notation.

\item The DDRSes for the binary view and the decimal view 
are hardly intelligible unless one knows that the objective is to construct a 
commutative ring. 
A decimal normal form is defined as either a 
digit, or an application of a decimal append function $\_\apd i$ to a nonzero normal form 
(for all digits $i$). 
This implies the absence of (superfluous) leading zeros, and the (ground) normal forms thus 
obtained correspond bijectively to the non-negative integers (that is, $\Nat$).
Incorporating all minus instances $-(t)$ of each nonzero normal form $t$ yields
the class of normal forms.
The ``semantics'' of these normal forms in the language of commutative rings is 
standard:
\begin{align*}
\llbracket 0\rrbracket&= 0,\\
\llbracket i'\rrbracket&=\llbracket i\rrbracket+1
  \quad\text{for all digits $0\leq i<9$ and $i'$ defined as in~Table~\ref{fig:enumdig},}\\
\llbracket x\apd i\rrbracket&=(\underline{10}\cdot \llbracket x\rrbracket)+\llbracket i\rrbracket
  \quad\text{for all digits $i$, and $\underline{10}=\llbracket 9\rrbracket+1$},\\
\llbracket -(x)\rrbracket&=-(\llbracket x\rrbracket).
\end{align*}
A binary normal form has similar semantics: 
$\llbracket x\apb i\rrbracket=(\underline{2}\cdot \llbracket x\rrbracket)+\llbracket i\rrbracket$
for  digits $0,1$, and $\underline{2}=1+1$.

\item Understanding the concept of a commutative ring can be expected only from a person who has
already acquired an understanding of the structure of integers and who accepts the concept of 
generalization of a structure to a class of structures sharing some but not all of its properties.

In other words, the understanding that a DDRS for the integers is provided in the binary view 
and in the decimal view can only
be communicated to an audience under the assumption that a reliable mental picture 
of the integers already exists in the minds of members of the audience. 
This mental picture, however, can 
in principle be communicated by taking notice of the DDRS for the unary view first.
This conceptual (near) circularity may be nevertheless be considered a significant 
weakness of the approach of defining (and even introducing) the
integers as an extension of naturals by means of rewriting.
\end{enumerate}
Although full confluence of the DDRSes defined in Section~\ref{sec:2} 
for the binary and decimal view
has been  disproven by \textsc{Kluiving} and \textsc{van Woer\-kom}~\cite{KW16} 
(with help of the confluence tool CSI~\cite{csi}), 
we prove in Appendix~\ref{app:A} that all DDRSes defined in this section
are ground-confluent, and thus ground-complete. 

In Section~\ref{sec:3} we discussed some alternatives for the above-mentioned DDRSes based on
papers of \textsc{Bouma} and \textsc{Walters}~\cite{BW89},
\textsc{Walters}~\cite{Wal94}, and \textsc{Walters} and \textsc{Zantema}~\cite{WZ95} 
in which digit tree constructors are used.
In~\cite{Wal94}, \textsc{Walters} presents a TRS (term rewriting system)
based on juxtaposition as a tree constructor for integer arithmetic with addition and subtraction 
that is ground-complete and parametric over any radix.
In~\cite{WZ95}, \textsc{Walters} and \textsc{Zantema} extend this TRS with multiplication 
and prove ground-completeness, using 
\emph{semantic labelling} for their termination proof, and judge this TRS | named JP | to have 
good efficiency 
and readability (in comparison with some alternatives discussed in that paper). 

With the tool AProVE~\cite{aprove}, \textsc{Kluiving} and \textsc{van Woerkom}~\cite{KW16}
proved strong termination of TRSes for arithmetic over $\Nat$ and over $\Int$ that employ 
the binary tree constructors.
In Appendix~\ref{app:B} we prove that the resulting DDRSes (Tables~\ref{tab:bc}
and~\ref{natbview}) are ground-complete.
\textsc{Kluiving} and \textsc{van Woerkom} also proposed in~\cite{KW16} a TRS for 
arithmetic over the natural numbers employing decimal
tree constructors (based on a DDRS proposed in 
version 2 of this paper) and proved strong termination with the tool AProVE. 
However, its natural extension to a TRS for integer arithmetic could not be proven terminating, 
probably due to its size.
This led to further research by \textsc{van der Kamp}~\cite{Kamp16}, who 
adapted both these TRSes and proved ground-completeness. The resulting DDRSes are
those in Tables~\ref{tab:natdecview} 
and~\ref{tab:intdecview}.

Of course, 
many normal forms in decimal notation have \emph{names} that confirm their base, 
for example
``six hundred eighty-nine'' $\langle\textsc{ae}\rangle$ or 
``six hundred and eighty-nine'' $\langle\textsc{be}\rangle$.
A decimal notation as 689 is so common that one usually does not question whether it 
represents $(6\apd 8)\apd 9$ or $(6\concd 8)\concd 9$ or some other formally defined notation.
Nevertheless, as we have seen, different algorithmic approaches to for example addition
may apply, although one would preferably not hamper an (initial) arithmetical method with notation 
such as $x\concd (y\concd z)$ and rewrite rules such as $x\concd (y\concd z)\to(x+ y)\concd z$, 
and for this reason we have a preference for
the DDRSes defined in Section~\ref{sec:2}. 
It should be noted that hand-written ground-confluence proofs of the size recorded in 
Appendix~\ref{app:A} and~\ref{app:B}
are of course error-prone and should be automated. 
Following~\cite{Kamp16}, we summarize in Table~\ref{tab:summa} 
the rule counts of the term rewriting systems for 
decimal representation of natural and integer arithmetic considered in this paper
and those considered in~\cite{WZ95} (the TRS named DA is discussed below), and observe
no significant differences in these counts.

\begin{table}
\hrule
~\\[2mm]
\centering
\begin{tabular}{lrrrrrrr}
Name
&\phantom{\hspace{3.2mm}}
$\Nat_{dub}$
&\phantom{\hspace{3.2mm}}
$\Int_{dub}$
&\phantom{\hspace{8.2mm}}
$\Nat_{dt}$
&\phantom{\hspace{3.2mm}}
$\Int_{dt}$
&\phantom{\hspace{8.2mm}}
DA(10)
&\phantom{\hspace{3.2mm}}
JP(10)
\\[2mm]
\hline
\\[-2mm]
rule count
&172
&444
&62
&277
&135
&438
\\[2mm]
rule schemes
&14
&32
&11
&28
&10
&30
\\[2mm]
extra operators
&$S(x)$,
&$S(x)$, $P(x)$,
&$S(x)$
&$S(x),$
&
&$x-y$
\\[1mm]
&$x\apb  0,~x\apb 1$
&$x\apb  0,~x\apb 1$
&
&$P(x)$
\\[2mm]
\end{tabular}
\hrule
\caption{Rule count for decimal representation, where 
DA(10) and JP(10) both originate from~\cite{WZ95} instantiated for base 10.} 
\label{tab:summa}
\end{table}

We briefly mention two other, comparable approaches to arithmetic that are also based on
some form of digit append constructors for representing  numbers. 
First, in~\cite{WZ95} \textsc{Walters} and \textsc{Zantema} introduce a TRS which they
named DA (for ``digit application'')
with addition and multiplication on natural numbers. The authors prove strong termination by 
recursive path ordering and confluence, and also judge this TRS to have 
good efficiency and readability. 
Secondly, in~\cite{CMR97}, \textsc{Contejean, March\'e} and \textsc{Rabehasaina} 
introduce integer arithmetic based on
\emph{balanced ternary numbers}, that is, numbers that can be represented by a digit append
function $\apt$ 
with digits -1,0,1 and semantics $\llbracket i\rrbracket= i$ and 
$\llbracket x\apt i\rrbracket= \underline3\cdot\llbracket x\rrbracket+i$ (see, e.g., 
\textsc{Knuth}~\cite{Knuth})
and provide a TRS that is confluent and terminating modulo associativity and commutativity of 
addition and multiplication.

Based on either a DDRS for the natural numbers or a DDRS for the integers one may develop a 
DDRS for rational numbers in various ways.
It is plausible to consider the meadow of rational numbers of~\cite{BergstraT07} or the
non-involutive meadow of rational numbers (see \cite{BM2014}) or the common meadow of rational 
numbers (see \cite{BP2014}) as abstract algebraic structures for rationals in which unary, 
binary, and decimal notation are to be incorporated in ways possibly based on the specifications 
presented above. 
Furthermore, one does well to consider the work discussed in~\cite{CMR97}
on a term rewriting system for rational numbers, in which
arithmetic for rational numbers is specified (this is the main result in~\cite{CMR97}, for which 
the above-mentioned work on integer arithmetic is a preliminary): the authors specify rational 
numbers by means of a TRS that is complete modulo associativity and commutativity of addition 
and multiplication, taking advantage of Stein's algorithm for computing gcd's of non-negative 
integers without any division%
  \footnote{Apart from halving even numbers, which is easy in binary notation, but
  can otherwise be specified with a shift operation.}
(see, e.g., \cite{Knuth}).

A survey of equational algebraic specifications for abstract datatypes is provided 
in~\cite{Wirsing91}.
In~\cite{BergstraT95} one finds the general result that computable abstract datatypes 
can be specified by means of
specifications which are confluent and strongly terminating term rewriting systems. 
Some general results on algebraic specifications can be found 
in~\cite{BroyWP84,BergstraT87,GaudelJ98}. 
More recent applications of equational specifications can be 
found in~\cite{BergstraT07}.

\bigskip

We conclude the paper with the introduction of a simple DDRS in Table~\ref{DDRSZ}
that specifies the integers in the signature $\Sigma_r=\{0,1,-(\_),+,\cdot\}$ of rings
and we name this datatype $\Int_r$.
Observe that the minus variant of equation~\ref{r7}, that is,
\[(-x)+(y+1)=((-x)+y)+1\]
is an instance of equation~\ref{r3}. 
\begin{table}
\hrule
\begin{minipage}[t]{0.5\linewidth}\centering
\begin{Lalign}
\label{r1}\tag*{[r1]}
-0&=0
\\
\label{r2}\tag*{[r2]}
-(-x) &= x
\\[3mm]
\label{r3}\tag*{[r3]}
x+(y+z) &= (x + y) + z
\\
\label{r4}\tag*{[r4]}
x+0&= x\\
\label{r5}\tag*{[r5]}
1+(-1)  &= 0 \\
\label{r6}\tag*{[r6]}
(x + 1)+(-1)  &= x \\
\label{r7}\tag*{[r7]}
x+(-(y+1))&=(x+(-y))+(-1)\\
\label{r8}\tag*{[r8]}
0+x     &= x\\
\label{r9}\tag*{[r9]}
(-1)+1  &= 0 \\
\label{r10}\tag*{[r10]}
(-(x+1)) + 1 &= -x\\
\label{r11}\tag*{[r11]}
(-x) + (-y) &= -(x+y)
\end{Lalign}
\end{minipage}\vspace{4mm}
\hfill
\begin{minipage}[t]{0.47\linewidth}\centering
\begin{Lalign}
\label{r12}\tag*{[r12]}
x \cdot 0 &= 0  \\
\label{r13}\tag*{[r13]}
x\cdot 1 &= x \\
\label{r14}\tag*{[r14]}
x\cdot (- y) &= (-x)\cdot y \\
\label{r15}\tag*{[r15]}
x\cdot(y + z) &= (x\cdot y) + (x\cdot z)
\end{Lalign}
\end{minipage}
\hrule
\caption{A DDRS for $\Int_r$, integer numbers in the language of rings}
\label{DDRSZ}
\end{table}
Also, observe that the equations in Table~\ref{DDRSZ} are semantic 
consequences of the axioms for commutative rings 
(equations $\eqref{e1} - \eqref{e8}$ in Table~\ref{tab:EqsForInt}).
In~\cite{KW16}, \textsc{Kluiving} and \textsc{van Woerkom} report that the term rewriting 
system defined by this DDRS
is strongly terminating,\footnote{Alternatively, the following weight function $|t|$
on closed terms can be used to prove strong termination: 
$|0| = |1| = 2, ~|-x|=2|x|+1, ~|x+y| = |x| + 3|y|$, and $|x\cdot y| = |x|\cdot|y|^2$.}
and below we prove that it is also ground-confluent, and thus ground-complete. 

Define the set $\NF$ of closed terms over $\Sigma_r$ as follows: 
\label{page:approach}
\begin{align*}
\NF&=\{0\}\cup\PNF\cup\NNF,\\
\PNF&= \{1\}\cup\{t+1\mid t\in\PNF\},\\
\NNF&=\{-t\mid t\in\PNF\}.
\end{align*}
It immediately follows that if $t\in\NF$, then $t$ is a normal form (no rewrite step
applies). Furthermore, two distinct 
elements in $\NF$ have distinct values in $\Int$. 
In order to prove ground-confluence of the
associated TRS it suffices to show that for each closed term over $\Sigma_r$,
either $t\in\NF$ or $t$ 
has a rewrite step, so that each normal form is in $\NF$.

We prove this by structural induction on $t$. The base cases $t\in\{0,1\}$ are
trivial. For the induction step we have to consider three cases:
\begin{enumerate}
\item
Case $t=-r$. Assume that $r\in\NF$ and apply case distinction on $r$:
\begin{itemize}
\item
if $r=0$, then $t\to 0$ by equation~\ref{r1},
\item
if $r\in\PNF$, then $t\in\NF$, 
\item 
if $r\in\NNF$, then $t$ has a rewrite step by equation~\ref{r2}.
\end{itemize}
\item
Case $t=u+r$. Assume that $u,r\in\NF$ and apply case distinction on $r$:
\begin{itemize}
\item
if $r=0$, then $t\to u$ by equation~\ref{r4},
\item
if $r=1$, then apply case distinction on $u$: 
\begin{itemize}
\item
if $u=0$, then $t\to1$ by equation~\ref{r8},
\item
if $u\in\PNF$, then $t\in\NF$,
\item 
if $u=-1$, then $t\to0$ by equation~\ref{r9},
\item
if $u=-(u'+1)$, then $t$ has a rewrite step by equation~\ref{r10},
\end{itemize}
\item 
if $r=r'+1$, then $t\to (u+r')+1$ by equation~\ref{r3},
\item 
if $r=-1$ then $t=u+(-1)$ and apply case distinction on $u$:
\begin{itemize}
\item
if $u=0$, then $t$ has a rewrite step by equation~\ref{r8},
\item
if $u=1$, then $t$ has a rewrite step by equation~\ref{r5},
\item
if $u=u'+1$, then $t$ has a rewrite step by equation \ref{r6},
\item
if $u\in\NNF$, then $t$ has a rewrite step by equation \ref{r11},
\end{itemize}
\item 
if $r= -(r'+1)$, then $t\to(u+(-r'))+(-1)$ by equation~\ref{r7}.
\end{itemize}

\item
Case $t=u\cdot r$. Assume that $u,r\in\NF$, then $t$ has a rewrite step according to
one of the equations $\text{\ref{r12}}-\text{\ref{r15}}$.
\end{enumerate}
This concludes our proof.

In~\cite{KW16}
it is observed that this DDRS for $\Int_r$ is not confluent:
$(-(-x))+(-y)\rightarrow x+(-y)$ and $(-(-x))+(-y)\rightarrow -((-x)+y)$ 
by~\ref{r2}, \ref{r11}. 
Attempts to use Knuth-Bendix completion yielded no solution and the authors write that
``Too many rules needed to be added and changed to solve the confluence issues. The system itself does not seem to be designed with confluence in mind''.

\paragraph{Acknowledgement.} We thank Boas Kluiving and Wijnand van Woerkom for 
adapting some of the DDRSes defined in version~2 of this paper and proving various results:
completeness of the DDRSes for $\Nat_{bud}$ and $\Nat_{dub}$;
strong termination and non-confluence of those for $\Int_{bud}$, $\Int_{dub}$, and $\Int_r$;
and reporting on all this in~\cite{KW16}. We thank Luca van der Kamp for his further research on  
DDRSes for $\Nat_{dt}$ and $\Int_{dt}$, and coming up 
with those defined in~\cite{Kamp16} and discussed in Section~\ref{sec:4.3}.

\appendix

\section{Ground-completeness proofs: DDRSes with digit append constructors}
\label{app:A}
In this appendix we prove ground-completeness for the DDRSes for $\Int_{ubd}$,
$\Int_{bud}$, and $\Int_{dub}$, respectively. 
In all ground-confluence proofs we adopt the approach used in that of the DDRS for
the ring of Integers (see page~\pageref{page:approach}).

\subsection{Unary view: the DDRS for $\Int_{ubd}$}
\label{app:A.1}
First we show that the term rewriting system defined
by the DDRS for $\Int_{ubd}$ in Table~\ref{tab:intunview} is strongly terminating. 
Define the following weight function $|t|$ on closed terms over $\Sigma_\Int$:
\begin{align*}
&|0|=2,&&|i'|=|i|+3~\text{ for }~i=0,\dots,8,\\
&|S(x)|=|x|+2,&&|x\apb i|=5\cdot|x|^7+1~\text{ for }~i=0,1,\\
&|P(x)|=|x|+5,&&|x\apd i|=29\cdot|x|^{31}+1~\text{ for }~i=0,1,\dots,9.\\
&|-x|=|x|+2,\\
&|x+y|=|x|\cdot|y|,\\
&|x\cdot y|=|x|^{|y|}.
\end{align*}
Then $|t|>1$ for all closed terms over $\Sigma_\Int$, and it easily follows that
each rewrite step on a closed term reduces its weight. (Of course, the defining equations
in the right column are superfluous: each closed term that matches one of its left-hand sides has 
a unique rewrite step to one that matches a left-hand side in the left column.)

Also, this rewriting systen is ground-confluent. Define the set $N$ as follows:
\begin{align*}
N&=\{0\}\cup N^+\cup N^-,\\
N^+&=\{S(0)\}\cup \{S(t)\mid t\in N^+\},\\
N^-&=\{-t\mid t\in N^+\}.
\end{align*}
It immediately follows that if $t\in N$, then $t$ is a normal form (no rewrite 
rule applies), and that two distinct elements in $N$ have distinct values in $\Int$.
Also, as stated in Section~\ref{sec:2.2}, the equations in Table~\ref{tab:intunview}
are semantic consequences of the axioms for commutative rings 
(equations $\eqref{e1} - \eqref{e8}$ in Table~\ref{tab:EqsForInt}).
In order to prove ground-confluence we have to show that
for each closed term $t$ over $\Sigma_\Int$, either $t\in N$ or $t$ has a rewrite step, 
so that each normal form is in $N$. 
We prove this by structural induction on $t$.

The base cases are simple: if $t=0$, then  $t\in N$, and 
if $t=i'$ for some $i\in\{0,1,2,3,4,5,6,7,8\}$, then $t\to S(i)$ by equation~\ref{u15}.

For the induction step we distinguish seven cases:
\begin{enumerate}
\item 
Case $t=S(r)$. Assume that $r\in N$ and apply case distinction on $r$:
\begin{itemize}
\item
if $r=0$, then $t\in N$,
\item
if $r=S(r')$ (thus $r'\in N^+$), then $t\in N$,
\item
if $r=-S(r')$, then $t\to -r'$ by equation~\ref{u6}.
\end{itemize}
\item
Case $t=P(r)$.
Assume that $r\in N$ and apply case distinction on $r$:
\begin{itemize}
\item
if $r=0$, then $t\to -S(0)$ by equation~\ref{u12},
\item
if $r=S(r')$, then $t\to r'$ by equation~\ref{u13},
\item
if $r=-S(r')$, then $t\to -S(S(r'))$ by equation~\ref{u14}.
\end{itemize}
\item 
Case $t=-r$. Assume that $r\in N$ and apply case distinction on $r$:
\begin{itemize}
\item
if $r=0$, then $t\to 0$ by equation~\ref{u5},
\item
if $r=S(r')$, then $t\in N$,
\item
if $r=-S(r')$, then $t\to S(r')$ by equation~\ref{u7}.
\end{itemize}
\item
Case $t=r\apb i$. Now $t$ has a rewrite step by equation~\ref{u16}.
\item
Case $t=r\apd i$.
Now
$t$ has a rewrite step by equation~\ref{u17}.
\item 
Case $t=u+r$. Assume that $u,r\in N$ and apply case distinction on $r$:
\begin{itemize}
\item
if $r=0$, then $t\to u$ by equation~\ref{u1},
\item
if $r=S(r')$, then $t\to S(u+r')$ by equation~\ref{u2},
\item
if $r=-S(r')$, then apply case distinction on $u$:
\begin{itemize}
\item
if $u=0$, then $t\to r$ by equation~\ref{u8},
\item
if $u=S(u')$, then $t\to S(u'+r)$ by equation~\ref{u9},
\item
if $u=-S(u')$, then $t\to -(S(u')+S(r'))$ by equation~\ref{u10}.
\end{itemize}
\end{itemize}
\item 
Case $t=u\cdot r$. Assume that $u,r\in N$ and apply case distinction on $r$:
\begin{itemize}
\item
if $r=0$, then $t\to 0$ by equation~\ref{u3},
\item
if $r=S(r')$, then $t\to (u\cdot r')+u$ by equation~\ref{u4},
\item
if $r=-S(r')$, then $t\to -(u\cdot r)$ by equation~\ref{u11}.
\end{itemize}
\end{enumerate}
This concludes our proof.

\subsection{Binary view: the DDRS for $\Int_{bud}$}
\label{app:A.2}
We prove that the term rewriting system defined by the DDRS for $\Int_{bud}$
in Table~\ref{tab:binview} is ground-complete.
This rewriting system is proven strongly terminating in~\cite{KW16}, so it remains 
to be proven that it is ground-confluent and again we adopt the approach used in the 
proof on page~\pageref{page:approach}.  

Define the set $N$ of closed terms over $\Sigma_\Int$ as follows:
\begin{align*}
N&=\{0\}\cup N^+\cup N^-,\\
N^+&=\{1\}\cup\{t\apb 0,t\apb 1\mid t\in N^+\},\\
N^-&=\{-t\mid t\in N^+\}.
\end{align*}
It immediately follows that if $t\in N$, then $t$ is a normal form (no rewrite 
rule applies), and that two distinct elements in $N$ have distinct values in $\Int$.
Also, as stated in Section~\ref{sec:2.2}, the equations in Tables~\ref{tab:natbinview}
and \ref{tab:binview} are semantic 
consequences of the axioms for commutative rings 
(equations $\eqref{e1} - \eqref{e8}$ in Table~\ref{tab:EqsForInt}).
In order to prove ground-confluence of this rewriting system we have to show that
for each closed term $t$ over $\Sigma_\Int$, either $t\in N$ or $t$ has a rewrite step, 
so that each normal form is in $N$. 
We prove this by structural induction on $t$.

The base cases are simple: if $t\in\{0,1\}$ then  $t\in N$, and 
if $t=i'$ for some $i\in\{1,2,3,4,5,6,7,8\}$, then $t\to S(i)$ by equation~\ref{b14}.

For the induction step we distinguish eight cases:
\begin{enumerate}

\item 
Case $t=S(r)$. Assume that $r\in N$ and apply case distinction on $r$:
\begin{itemize}
\item
if $r=0$, then $t\to 1$ by equation~\ref{b2},
\item
if $r=1$, then $t\to 1\apb 0$ by equation~\ref{b3},
\item
if $r=r'\apb 0$, then $t\to r'\apb 1$ by equation~\ref{b4},
\item
if $r=r'\apb 1$, then $t\to S(r')\apb 0$ by equation~\ref{b5},
\item
if $r=-1$, then $t\to 0$ by equation~\ref{b23},
\item
if $r=-(r'\apb 0)$, then $t\to -(P(r')\apb 1)$ by equation~\ref{b24},
\item 
if $r=-(r'\apb 1)$, then $t\to -(r'\apb 0)$ by equation~\ref{b25}.
\end{itemize}
\item
Case $t=P(r)$. Assume that $r\in N$ and apply case distinction on $r$:
\begin{itemize}
\item
if $r=0$, then $t\to -1$ by equation~\ref{b18},
\item
if $r=1$, then $t\to 0$ by equation~\ref{b19},
\item
if $r=r'\apb 0$, then $t\to P(r')\apb 1$ by equation~\ref{b20},
\item
if $r=r'\apb 1$, then $t\to r'\apb 0$ by equation~\ref{b21},
\item
if $r=-1$, then $t\to -S(1)$ by equation~\ref{b22},
\item
if $r=-(r'\apb i)$, then $t\to -S(r'\apb i)$ by equation~\ref{b22}.
\end{itemize}
\item 
Case $t=-r$. Assume that $r\in N$ and apply case distinction on $r$:
\begin{itemize}
\item
if $r=0$, then $t\to 0$ by equation~\ref{b16},
\item
if $r=1$, then $t\in N$,
\item
if $r=r'\apb i$, then $t\in N$,
\item
if $r=-1$, then $t\to 1$ by equation~\ref{b17},
\item
if $r=-(r'\apb i)$, then $t\to r'\apb i$ by equation~\ref{b17}.
\end{itemize}
\item
Case $t=r\apb 0$. Assume that $r\in N$ and apply case distinction on $r$:
\begin{itemize}
\item
if $r=0$, then $t\to 0$ by the first equation of~\ref{b1},
\item
if $r=1$ or $r=r'\apb i$, then $t\in N$,
\item
if $r=-1$ or $r=-(r'\apb i)$, then $t$ has a rewrite step by equation~\ref{b26}.
\end{itemize}
\item
Case $t=r\apb 1$. Assume that $r\in N$ and apply case distinction on $r$:
\begin{itemize}
\item
if $r=0$, then $t\to j$ by the second equation of~\ref{b1},
\item
if $r=1$ or $r=r'\apb i$, then $t\in N$,
\item
if $r=-1$ or $r=-(r'\apb i)$, then $t$ has a rewrite step by equation~\ref{b27}.
\end{itemize}
\item
Case $t=r\apd i$.
Now
$t$ has a rewrite step by equation~\ref{b15}.
\item 
Case $t=u+r$. Assume that $u,r\in N$ and apply case distinction on $r$:
\begin{itemize}
\item
if $r=0$, then $t\to u$ by equation~\ref{b6},
\item
if $r=1$, then $t\to S(u)$ by equation~\ref{b8},
\item 
if $r=r'\apb i$, apply case distinction on $u$:
\begin{itemize}
\item
if $u=0$, then $t\to r$ by equation~\ref{b7},
\item
if $u=1$, then $t\to S(r)$ by equation~\ref{b9},
\item
if $u=u'\apb j$, then $t$ has a rewrite step according to one of~\ref{b10},
\item
if $u=-1$, then $t\to P(r)$ by equation~\ref{b29},
\item
if $u=-(u'\apb j)$, then $t$ has a rewrite step according to one of~\ref{b31},
\end{itemize}
\item
if $r=-1$, then $t\to P(u)$ by equation~\ref{b28},
\item 
if $r=-(r'\apb i)$, apply case distinction on $u$:
\begin{itemize}
\item
if $u=0$, then $t\to r$ by equation~\ref{b7},
\item
if $u=1$, then $t\to S(r)$ by equation~\ref{b9},
\item
if $u=u'\apb j$, then $t$ has a rewrite step according to one of~\ref{b30},
\item
if $u=-1$, then $t\to P(r)$ by equation~\ref{b29},
\item
if $u=-(u'\apb j)$, then $t$ has a rewrite step by equation~\ref{b32}.
\end{itemize}
\end{itemize}
\item 
Case $t=u\cdot r$. Assume that $u,r\in N$ and apply case distinction on $r$:
\begin{itemize}
\item
if $r=0$, then $t\to 0$ by equation~\ref{b11},
\item
if $r=1$, then $t\to u$ by equation~\ref{b12},
\item 
if $r=r'\apb i$, then $t$ has a rewrite step according to one of~\ref{b13},
\item
if $r=-1$ or $r=-(r'\apb i)$, then $t$ has a rewrite step by equation~\ref{b33}.
\end{itemize}
\end{enumerate}
This concludes our proof.

\subsection{Decimal view: the DDRS for $\Int_{dub}$}
\label{app:A.3}
We prove that the term rewriting system defined by the DDRS for $\Int_{bud}$
in Table~\ref{tab:indecview} 
(using $i^\star$ as defined in Table~\ref{fig:subdig})
is ground-complete.
This rewriting system is proven strongly terminating in~\cite{KW16}, so it remains 
to be proven that it is ground-confluent.  

Recall we write $D$  for the set of all digits.  
Define the set $N$ of closed terms over $\Sigma_\Int$ as follows:
\begin{align*}
N&=\{0\}\cup N^+\cup N^-,\\
N^+&=D\setminus\{0\}\cup\{t\apb i\mid t\in N^+, i\in D\},\\
N^-&=\{-t\mid t\in N^+\}.
\end{align*}
It immediately follows that if $t\in N$, then $t$ is a normal form (no rewrite 
rule applies), and that two distinct elements in $N$ have distinct values in $\Int$.
Also, as stated in Section~\ref{sec:2.3}, the equations in Tables~\ref{tab:decview}
and \ref{tab:indecview} are semantic 
consequences of the axioms for commutative rings 
(equations $\eqref{e1} - \eqref{e8}$ in Table~\ref{tab:EqsForInt}).
In order to prove ground-confluence of this rewriting system we have to show that
for each closed term $t$ over $\Sigma_\Int$, either $t\in N$ or $t$ has a rewrite step, 
so that each normal form is in $N$. 
We prove this by structural induction on $t$.

The base cases are trivial: if $t\in D$, then  $t\in N$.

For the induction step we distinguish eight cases:
\begin{enumerate}

\item 
Case $t=S(r)$. Assume that $r\in N$ and apply case distinction on $r$:
\begin{itemize}
\item
if $r=i$ for $i\in\{0,1,\dots,8\}$, then $t\to i'$ by equation~\ref{d2},
\item
if $r=9$, then $t\to 1\apd 0$ by equation~\ref{d3},
\item
if $r=r'\apd i$ for $i\in\{0,1,\dots,8\}$, then $t\to r'\apd i'$ by equation~\ref{d4},
\item
if $r=r'\apd 9$, then $t\to S(r')\apd 0$ by equation~\ref{d5},
\item
if $r=-i'$ for $i\in\{0,1,\dots,8\}$, then $t\to -i$ by equation~\ref{d22},
\item
if $r=-(r'\apd 0)$, then $t\to -(P(r')\apd 9)$ by equation~\ref{d23},
\item 
if $r=-(r'\apd i')$ for $i\in\{0,1,\dots,8\}$, then $t\to -(r'\apd i)$ by equation~\ref{d24}.
\end{itemize}
\item
Case $t=P(r)$. Assume that $r\in N$ and apply case distinction on $r$:
\begin{itemize}
\item
if $r=0$, then $t\to -1$ by equation~\ref{d17},
\item
if $r=i'$ for $i\in\{0,1,\dots,8\}$, then $t\to i$ by equation~\ref{d18},
\item
if $r=r'\apd 0$, then $t\to P(r')\apd 9$ by equation~\ref{d19},
\item
if $r=r'\apd i'$ for $i\in\{0,1,\dots,8\}$, then $t\to r'\apd i$ by equation~\ref{d20},
\item
if $r=-i'$ for $i\in\{0,1,\dots,8\}$, then $t\to -S(i')$ by equation~\ref{d21},
\item
if $r=-(r'\apd i)$, then $t\to -S(r'\apd i)$ by equation~\ref{d21}.
\end{itemize}
\item 
Case $t=-r$. Assume that $r\in N$ and apply case distinction on $r$:
\begin{itemize}
\item
if $r=0$, then $t\to 0$ by equation~\ref{d15},
\item
if $r=i'$ for $i\in\{0,1,\dots,8\}$, then $t\in N$,
\item
if $r=r'\apd i$, then $t\in N$,
\item
if $r=-i'$ for $i\in\{0,1,\dots,8\}$, then $t\to i'$ by equation~\ref{d16},
\item
if $r=-(r'\apd i)$, then $t\to r'\apd i$ by equation~\ref{d16}.
\end{itemize}

\item
Case $t=r\apd 0$. Assume that $r\in N$ and apply case distinction on $r$:
\begin{itemize}
\item
if $r=0$, then $t\to 0$ by the first equation of~\ref{d1},
\item
if $r=i'$ for $i\in\{0,1,\dots,8\}$, then $t\in N$,
\item 
if $r=r'\apd i$, then $t\in N$,
\item
if $r\in N^-$, then $t\to -(r\apd 0)$ by equation~\ref{d25}.
\end{itemize}
\item
Case $t=r\apd j$ for $j\in\{1,2,\dots,9\}$. Assume that $r\in N$ and apply case distinction on $r$:
\begin{itemize}
\item
if $r=0$, then $t\to j$ by the appropriate equation of~\ref{d1},
\item
if $r=i'$ for $i\in\{0,1,\dots,8\}$, then $t\in N$,
\item 
if $r=r'\apd i$, then $t\in N$,
\item
if $r\in N^-$, then $t\to -(P(r)\apd j^\star)$ by one of the equations of~\ref{d26}.
\end{itemize}
\item
Case $t=r\apb i$.
Now
$t$ has a rewrite step by equation~\ref{d14}.
\item 
Case $t=u+r$. Assume that $u,r\in N$ and apply case distinction on $r$:
\begin{itemize}
\item
if $r=0$, then $t\to u$ by equation~\ref{d6},
\item
if $r=i$ for $i\in\{1,2,\dots,9\}$, then $t\to S^i(u)$ by equation~\ref{d8},
\item 
if $r=r'\apd i$, apply case distinction on $u$:
\begin{itemize}
\item
if $u=0$, then $t\to r$ by equation~\ref{d7},
\item
if $u=j$ for $j\in\{1,2,\dots,9\}$, then $t\to S^j(r)$ by equation~\ref{d9},
\item
if $u=u'\apd j$, then $t$ has a rewrite step according to one of~\ref{d10},
\item
if $u=-j$ for $j\in\{1,2,\dots,9\}$, then $t\to P^j(r)$ by one of~\ref{d28},
\item
if $u=-(u'\apd j)$, then $t$ has a rewrite step according to one of~\ref{d30},
\end{itemize}
\item
if $r=-i$ for $i\in\{1,2,\dots,9\}$, then $t\to P^i(u)$ by one of the equations of~\ref{d27},
\item 
if $r=-(r'\apd i)$, apply case distinction on $u$:
\begin{itemize}
\item
if $u=0$, then $t\to r$ by equation~\ref{d7},
\item
if $u=j$ for $j\in\{1,2,\dots,9\}$, then $t\to S^j(r)$ by one of~\ref{d9},
\item
if $u=u'\apd j$, then $t$ has a rewrite step according to one of~\ref{d29},
\item
if $u=-j$ for $j\in\{1,2,\dots,9\}$, then $t\to P^j(r)$ by equation~\ref{d28},
\item
if $u=-(u'\apd j)$, then $t$ has a rewrite step by equation~\ref{d31}.
\end{itemize}
\end{itemize}
\item 
Case $t=u\cdot r$. Assume that $u,r\in N$ and apply case distinction on $r$:
\begin{itemize}
\item
if $r=0$, then $t\to 0$ by equation~\ref{d11},
\item
if $r=i'$ for $i\in\{0,1,\dots,8\}$, then $t\to (u\cdot i)+u$ by equation~\ref{d12},
\item
if $r=r'\apb i$ for $i\in\{0,1,\dots,9\}$, then $t\to((u\cdot r')\apd 0)+(u\cdot i)$ 
by equation~\ref{d13},
\item 
if $r\in N^-$, then $t$ has a rewrite step by equation~\ref{d32}.
\end{itemize}
\end{enumerate}
This concludes our proof.

\section{Ground-completeness proofs: DDRSes with digit tree constructors}
\label{app:B}
In Appendix~\ref{app:B.1} we prove ground-completeness of the DDRS for $\Int_{ut}$, and
in Appendix~\ref{app:B.2} we prove ground-completeness of the DDRS for $\Int_{bt}$.

\subsection{Unary view: the DDRS for $\Int_{ut}$}
\label{app:B.1}
First we show that the term rewriting system defined by the DDRS for $\Int_{ut}$ in 
Table~\ref{tab:intuaview} is strongly terminating. 

Define the signature $\Sigma_{ut}=\{0,-(\_),~\_\concu\_~,+,\cdot\}$ and
the following weight function $|t|$ on closed terms over $\Sigma_{ut}$:
\begin{align*}
&|0|=1,\\
&|-x|=|x|+1,\\
&|x\concu y|=|x|+2|y|,\\
&|x+y|=|x|+2|y|,\\
&|x\cdot y|=2\cdot|x|\cdot|y|.
\end{align*}
Then $|t|\geq1$ for all closed terms over $\Sigma_{ut}$, and it easily follows that
each rewrite step on a closed term reduces its weight.

Also, this rewriting system is ground-confluent. Define the set $N$ as follows:
\begin{align*}
N&=\{0\}\cup N^+\cup N^-,\\
N^+&=\{0\concu 0\}\cup \{t\concu 0\mid t\in N^+\},\\
N^-&=\{-t\mid t\in N^+\}.
\end{align*}
It immediately follows that if $t\in N$, then $t$ is a normal form (no rewrite 
rule applies), and that two distinct elements in $N$ have distinct values in $\Int$.
Also, as stated in Section~\ref{subsec:3.1}, the equations in Table~\ref{tab:intuaview}
are semantic consequences of the axioms for commutative rings 
(equations $\eqref{e1} - \eqref{e8}$ in Table~\ref{tab:EqsForInt}).
In order to prove ground-confluence of the DDRS for $\Int_{ut}$ we have to show that
for each closed term $t$ over $\Sigma_{ut}$, either $t\in N$ or $t$ has a rewrite step, 
so that each normal form is in $N$. 
We prove this by structural induction on $t$.

The base case is trivial: if $t=0$, then  $t\in N$.

For the induction step we distinguish four cases:
\begin{enumerate}
\item 
Case $t=-r$. Assume that $r\in N$ and apply case distinction on $r$:
\begin{itemize}
\item
if $r=0$, then $t\to 0$ by equation~\ref{ut6},
\item
if $r=r'\concu 0$, then $t\in N$,
\item 
if $r=-(r'\concu 0)$, then $t\to r'\concu 0$ by equation~\ref{ut7}.
\end{itemize}
\item
Case $t=v\concu r$. Assume that $v,r\in N$ and apply case distinction on $r$:
\begin{itemize}
\item
if $r=0$, then apply case distinction on $v$:
\begin{itemize}
\item
if $v=0$, then $t\in N$,
\item
if $v=v'\concu 0$, then $t\in N$,
\item
if $v=-(v'\concu 0)$, then $t\to -v'$ by equation~\ref{ut10}.
\end{itemize}
\item 
if $r=r'\concu 0$, then apply case distinction on $v$:
\begin{itemize}
\item
if $v=0$, then $t\to (0\concu r')\concu 0$ by equation~\ref{ut1},
\item
if $v=v'\concu 0$, then $t\to (v\concu r')\concu 0$ by equation~\ref{ut1},
\item
if $v=-(v'\concu 0)$, then $t\to (-v')\concu r'$ by equation~\ref{ut11}.
\end{itemize}
\item 
if $r=-(r'\concu 0)$, then apply case distinction on $v$:
\begin{itemize}
\item
if $v=0$, then $t\to -r'$ by equation~\ref{ut8},
\item
if $v=v'\concu 0$, then $t\to v'\concu(-r')$ by equation~\ref{ut9},
\item 
if $v=-(v'\concu 0)$, then $t\to ((v'+r')\concu 0)$ by equation~\ref{ut12}.
\end{itemize}
\end{itemize}
\item 
Case $t=v+r$. Assume that $v,r\in N$ and apply case distinction on $r$:
\begin{itemize}
\item
if $r=0$, then $t\to v$ by equation~\ref{ut2},
\item
if $r=r'\concu 0$, then $t\to (v+r')\concu 0$ by equation~\ref{ut3},
\item
if $r=-(r'\concu 0)$, then apply case distinction on $v$:
\begin{itemize}
\item
if $v=0$, then $t\to r'$ by equation~\ref{ut13},
\item
if $v=v'\concu 0$, then $t\to v'+(-r')$ by equation~\ref{ut14},
\item
if $v=-(v'\concu 0)$, then $t\to -((v'\concu 0)+(r'\concu 0))$ by equation~\ref{ut15}.
\end{itemize}
\end{itemize}
\item 
Case $t=v\cdot r$. Assume that $v,r\in N$ and apply case distinction on $r$:
\begin{itemize}
\item
if $r=0$, then $t\to 0$ by equation~\ref{ut4},
\item
if $r=r'\concu 0$, then $t\to (v\cdot r')+v$ by equation~\ref{ut5},
\item
if $r=-(r'\concu 0)$, then $t\to -(v\cdot r)$ by equation~\ref{ut16}.
\end{itemize}
\end{enumerate}
This concludes our proof.

\subsection{Binary view: the DDRS for $\Int_{bt}$}
\label{app:B.2}
We prove that the term rewriting system defined by the DDRS for $\Int_{bt}$
in Table~\ref{natbview} is ground-complete.
This rewriting system is proven strongly terminating in~\cite{KW16}, so it remains 
to be proven that this DDRS is ground-confluent, and we adopt the approach used in the 
proof on page \pageref{page:approach}.  

Define the signature $\Sigma_{bt}=\{0,1,-(\_),~\_\concb\_~,+,\cdot\}$, and
the set $N$ of closed terms over $\Sigma_{bt}$ as follows:
\begin{align*}
N&=\{0\}\cup N^+\cup N^-,\\
N^+&=\{1\}\cup\{t\concb 0,t\concb 1\mid t\in N^+\},\\
N^-&=\{-t\mid t\in N^+\}.
\end{align*}
It immediately follows that if $t\in N$, then $t$ is a normal form (no rewrite 
rule applies), and that two distinct elements in $N$ have distinct values in $\Int$.
Observe that the equations in Table~\ref{natbview} are semantic 
consequences of the axioms for commutative rings 
(equations $\eqref{e1} - \eqref{e8}$ in Table~\ref{tab:EqsForInt}).
In order to prove ground-confluence of this rewriting system we have to show that
for each closed term $t$ over $\Sigma_{bt}$, either $t\in N$ or $t$ has a rewrite step, 
so that each normal form is in $N$. 
We prove this by structural induction on $t$.

The base cases are simple: if $t\in\{0,1\}$, then  $t\in N$.

For the induction step we distinguish four cases:
\begin{enumerate}
\item 
Case $t=-r$. Assume that $r\in N$ and apply case distinction on $r$:
\begin{itemize}
\item
if $r=0$, then $t\to 0$ by equation~\ref{bi13},
\item
if $r=1$, then $t\in N$,
\item
if $r=r'\concb i$, then $t\in N$,
\item
if $r=-1$, then $t\to 1$ by equation~\ref{bi14},
\item
if $r=-(r'\concb i)$, then $t\to r'\concb i$ by equation~\ref{bi14}.
\end{itemize}
\item
Case $t=r\concb u$. Assume that $r,u\in N$ and apply case distinction on $r$:
\begin{itemize}
\item
if $r=0$, then $t\to u$ by equation~\ref{bi1},
\item
if $r=1$ apply case distinction on $u$:
\begin{itemize}
\item
if $u=0$, then $t\in N$,
\item
if $u=1$, then $t\in N$,
\item
if $u=u'\concb i$, then $t\to (1+u')\concb i$ by equation~\ref{bi2},
\item
if $u=-1$, then $t\to 1$ by equation~\ref{bi15},
\item
if $u=-(u'\concb j)$, then $t\to -((u'+(-1))\concb j)$ by equation~\ref{bi18},
\end{itemize}
\item
if $r=r'\concb i$, then apply case distinction on $u$:
\begin{itemize}
\item
if $u\in\{0,1\}$, then $t\in N$,
\item
if $u=u'\concb j$, then $t\to (r+u')\concb j$ by equation~\ref{bi2},
\item
if $u=-1$ and $i=0$, then $t\to (r'\concb (-1))\concb 1$ by equation~\ref{bi16},
\item
if $u=-1$ and $i=1$, then $t\to (r'\concb 0)\concb 1$ by equation~\ref{bi17},
\item
if $u=-(u'\concb j)$, then $t\to -((u'+ (-r))\concb j)$ by equation~\ref{bi18},
\end{itemize}
\item
if $r=-1$, then apply case distinction on $u$:
\begin{itemize}
\item
if $u\in\{0,1,-1\}$, then $t\to-(1\concb (-u))$ by equation~\ref{bi19},
\item
if $u=u'\concb j$, then $t\to((-1)+u')\concb j$ by equation~\ref{bi2},
\item
if $u=-(u'\concb j)$, then $t\to -((u'+(-(-1)))\concb j)$ by equation~\ref{bi18}, 
\end{itemize}
\item
if $r=-(r'\concb i)$, then $t$ has a rewrite step by equation~\ref{bi19}.
\end{itemize}

\item 
Case $t=u+r$. Assume that $u,r\in N$ and apply case distinction on $r$:
\begin{itemize}
\item
if $r=0$, then $t\to u$ by equation~\ref{bi4},
\item
if $r=1$, then apply case distinction on $u$:
\begin{itemize}
\item
if $u=0$, then $t\to1$ by equation~\ref{bi3},
\item
if $u=1$, then $t\to(1\concb 0)$ by equation~\ref{bi5}
\item
if $u=u'\concb j$ then $t\to u'\concb (j+1)$ by equation~\ref{bi7},
\item
if $u=-1$, then $t\to0$ by equation~\ref{bi21},
\item
if $u=-(u'\concb j)$ then $t\to -(u'\concb (j+(-1)))$ by equation~\ref{bi24},
\end{itemize}
\item 
if $r=r'\concb i$, then $t\to r'\concb (u+j)$ by \ref{bi6},
\item
if $r=-1$, then apply case distinction on $u$:
\begin{itemize}
\item
if $u=0$, then $t\to r$ by equation~\ref{bi3},
\item
if $u=1$, then $t\to0$ by equation~\ref{bi20}
\item
if $u=u'\concb j$ then $t\to u'\concb (j+(-1))$ by equation~\ref{bi7},
\item
if $u=-1$, then $t\to -(1\concb 0)$ by equation~\ref{bi22},
\item
if $u=-(u'\concb j)$ then $t\to -(u'\concb(j+(-r))$ by equation~\ref{bi24}, 
\end{itemize}
\item 
if $r=-(r'\concb i)$, then $t\to -(r'\concb (i+(-u)))$ by equation~\ref{bi23}. 
\end{itemize}

\item 
Case $t=u\cdot r$. Assume that $u,r\in N$ and apply case distinction on $r$:
\begin{itemize}
\item
if $r=0$, then $t\to 0$ by equation~\ref{bi8},
\item
if $r=1$, then apply case distinction on $u$:
\begin{itemize}
\item
if $u=0$, then $t\to 0$ by equation\ref{bi9},
\item
if $u=1$, then $t\to 1$ by equation\ref{bi10},
\item
if $u=u'\concb j$, then $t\to (u'\cdot r)\concb (j\cdot r)$ by equation\ref{bi12},
\item
if $u\in N^-$, then $t$ has a rewrite step by equation~\ref{bi26},
\end{itemize}
\item 
if $r=r'\concb i$, then $t\to(u\cdot r')\concb(u\cdot i)$ by equation~\ref{bi11},
\item
if $r\in N^-$, then $t$ has a rewrite step by equation~\ref{bi25}.
\end{itemize}
\end{enumerate}
This concludes our proof.

\section{Another unary view: unary append}
\label{app:C}
We briefly consider a simple alternative notation for the unary view that is related to 
tallying and establishes a unary numeral system based on unary digit append.
However, using only one digit requires this digit to be 0 for the representation of zero,
while the semantics of the ``unary digit append'' function $\_\apu0$ requires the appended 
digit 0 to have value 1, that is
\[\llbracket t\apu0\rrbracket = \llbracket t \rrbracket +1.\]
Note that this mismatch does not occur in our numeral system for unary view with digit tree
constructor in Section~\ref{subsec:3.1}, because in that case application of the constructor
function $\_\concu\_$ does not refer to 0 as a value.

In order to solve this mismatch, we introduce the one-place function (postfix)
\[\_\apue: \Int \to \Int,
\quad\text{the \emph{unary append},}
\] 
and define the datatypes $\Nat_{u'}$ and $\Int_{u'}$ based on the constant 0 and 
unary append. 
Consider the signature 
$\Sigma_{u'}=\{0, -(\_),~\_\apue, +, \cdot\}$.
In Table~\ref{tab:natuview} we define a DDRS for the datatype $\Nat_{u'}$ over
$\Sigma_{u'}$.
Of course, the phenomenon of ``removing leading zeros'' does not exist in this particular unary 
view.
Normal forms are 0 for zero, and applications of the unary 
append function that define all successor values: each natural
number $n$ is represented by $n$ applications of the unary append to 0 and can be seen
as representing a sequence of $1$'s of length $n$ having 0 as a single prefix, e.g.
\[(0\apue)\apue\]
is the normal form that represents $2$.

\begin{table}
\centering
\hrule
\begin{minipage}[t]{0.4\linewidth}\centering
\begin{Lalign}
\label{z1}
\tag*{[u$'$1]}
x+0 &=x\\
\label{z2}
\tag*{[u$'$2]}
x + (y\apue) &= (x\apue)+y
\end{Lalign}
\end{minipage}
\hspace{8mm}
\begin{minipage}[t]{0.5\linewidth}\centering
\begin{Lalign}
\label{z3}
\tag*{[u$'$3]}
x \cdot 0 &=0\\
\label{z4}
\tag*{[u$'$4]}
x \cdot (y\apue) &= (x\cdot y)+x
\end{Lalign}
\end{minipage}\vspace{4mm}
\hrule
\caption{A DDRS for~$\Nat_{u'}$, natural numbers in unary view with zero append}
\label{tab:natuview}
\end{table}

The transition to integer numbers is straightforward.
All minus instances $-t$ of nonzero normal forms $t$ define the negative normal forms, e.g.
\[-((0\apue)\apue)\]
is the normal form that represents $-2$. 
A DDRS that defines the extension of $\Nat_{u'}$ to integer numbers $\Int_{u'}$ is given in 
Table~\ref{tab:intuview}.

\begin{table}[b]
\centering
\hrule
\begin{minipage}[t]{0.4\linewidth}\centering
\begin{Lalign}
\tag*{[u$'$1]}
x+0 &=x\\
\tag*{[u$'$2]}
x + (y\apue) &= (x\apue)+y
\\[2mm]
\tag*{[u$'$3]}
x \cdot 0 &=0\\
\tag*{[u$'$4]}
x \cdot (y\apue) &= (x\cdot y)+x
\end{Lalign}
\end{minipage}
\hspace{8mm}
\begin{minipage}[t]{0.50\linewidth}\centering
\begin{Lalign}
\label{z5}
\tag*{[u$'$5]}
-0&=0\\
\label{z6}
\tag*{[u$'$6]}
(-(x\apue))\apue&=-x\\
\label{z7}
\tag*{[u$'$7]}
-(-x) &=x
\\[2mm]
\label{z8}
\tag*{[u$'$8]}
0+x&=x\\
\label{z9}
\tag*{[u$'$9]}
(x\apue)+(-(y\apue))&=x+(-y)
\\
\label{z10}
\tag*{[u$'$10]}
(-x)+(-y)&=-(x+y)
\\[2mm]
\label{z11}
\tag*{[u$'$11]}
x \cdot (-y) &=-(x\cdot y)
\end{Lalign}
\end{minipage}\vspace{4mm}
\hrule
\caption{A DDRS for~$\Int_{u'}$ that specifies
integer numbers in unary view with zero append}
\label{tab:intuview}
\end{table}

Define the following weight function $|t|$ on closed terms over $\Sigma_{u'}$:
\begin{align*}
&|0|=1,\\
&|-x|=|x|+1,\\
&|x\apue|=|x|+2,\\
&|x+y|=|x|+2|y|,\\
&|x\cdot y|=2\cdot|x|\cdot|y|.
\end{align*}
Then $|t|>0$ for each closed term $t$ and it easily follows that
each rewrite step on a closed term reduces its weight. Hence, both these DDRSes for 
$\Nat_{u'}$ and $\Int_{u'}$ define
a strongly terminating rewriting system.

We now prove that both these DDRSes are ground-confluent. We prove this for the latter, which 
implies ground-confluence
of the former.
Define the set $N$ as follows:
\begin{align*}
N&=\{0\}\cup N^+\cup N^-,\\
N^+&=\{0\apue\}\cup \{t\apue\mid t\in N^+\},\\
N^-&=\{-t\mid t\in N^+\}.
\end{align*}
It immediately follows that if $t\in N$, then $t$ is a normal form (no rewrite 
rule applies), and that two distinct elements in $N$ have distinct values in $\Int$.
Also, the equations in Table~\ref{tab:intuview}
are semantic consequences of the axioms for commutative rings 
(equations $\eqref{e1} - \eqref{e8}$ in Table~\ref{tab:EqsForInt}).
In order to prove ground-confluence we have to show that
for each closed term $t$ over $\Sigma_{u'}$, either $t\in N$ or $t$ has a rewrite step, 
so that each normal form is in $N$. 
We prove this by structural induction on $t$.

The base case is simple: if $t=0$, then  $t\in N$.

For the induction step we have to distinguish four cases:
\begin{enumerate}
\item 
Case $t=-r$. Assume that $r\in N$ and apply case distinction on $r$:
\begin{itemize}
\item
if $r=0$, then $t\to 0$ by equation~\ref{z5},
\item
if $r=r'\apue$, then $t\in N$,
\item
if $r=-(r'\apue)$, then $t\to r'\apue$ by equation~\ref{z7}.
\end{itemize}
\item
Case $t=r\apue$. Assume that $r\in N$ and apply case distinction on $r$:
\begin{itemize}
\item
if $r=0$, then $t\in N$,
\item
if $r=r'\apue$, then $t\in N$,
\item
if $r=-(r'\apue)$, then $t\to -r'$ by equation~\ref{z6}.
\end{itemize}
\item 
Case $t=u+r$. Assume that $u,r\in N$ and apply case distinction on $r$:
\begin{itemize}
\item
if $r=0$, then $t\to u$ by equation~\ref{z1},
\item
if $r=r'\apue$, then $t\to (u\apue) + r'$ by equation~\ref{z2},
\item
if $r=-(r'\apue)$, then apply case distinction on $u$:
\begin{itemize}
\item
if $u=0$, then $t\to r$ by equation~\ref{z8},
\item
if $u=u'\apue$, then $t\to u'+-(r')$ by equation~\ref{z9},
\item
if $u=-(u'\apue)$, then $t\to -((u'\apue)+(r'\apue))$ by equation~\ref{z10}.
\end{itemize}
\end{itemize}
\item 
Case $t=u\cdot r$. Assume that $u,r\in N$ and apply case distinction on $r$:
\begin{itemize}
\item
if $r=0$, then $t\to 0$ by equation~\ref{z3},
\item
if $r=r'\apue$, then $t\to (u\cdot r')+u$ by equation~\ref{z4},
\item
if $r=-(r'\apue)$, then $t\to -(u\cdot (r'\apue))$ by equation~\ref{z11}.
\end{itemize}
\end{enumerate}
This concludes our proof.

\bigskip

Finally, we observe that writing $S(\_)$ for the unary append function $\_\apue$ yields 
alternative specifications of equal size
for the datatypes $\Nat_{ubd}$ and $\Int_{ubd}$ defined in Section~\ref{subsec:2.1}
(Tables~\ref{tab:natunview} and~\ref{tab:intunview}) when disregarding their
equations for predecessor%
  \footnote{Observe that $\Int_{ubd}$ could have been specified without the predecessor $P(x)$
  and its three defining equations.} 
and for binary and decimal notation. 
Tagging the resulting equations with [u$''n$], we observe that addition on the naturals is  
defined by the two equations
\begin{Lalign}
\tag*{[u$''$1]}
x+0 &=x\\
\tag*{[u$''$2]}
x + S(y) &= S(x)+y\hspace{10cm}
\end{Lalign} 
where [u$''$2] (as a rewrite rule) is not standard, while the extra equations used to define 
addition on the integers,
\begin{Lalign}
\tag*{[u$''$8]}
0+x &=x\\
\tag*{[u$''$9]}
S(x) + (-S(y)) &= x+(-y)\\
\tag*{[u$''$10]}
(-x) + (-y) &= -(x+y)\hspace{9cm}
\end{Lalign} 
are quite natural (only [u$''$9] differs from~\ref{u9} in Table~\ref{tab:intunview}).
The remaining six equations for $\Int_{u'}$ in this notation
define multiplication and minus, and exactly match those defined for $\Int_{ubd}$.
\end{document}